\documentclass[11pt]{article}
\usepackage[letterpaper,margin=1in]{geometry}
\usepackage[T1]{fontenc}
\usepackage[utf8]{inputenc}
\usepackage{lmodern}
\usepackage{microtype}
\usepackage{amsmath,amssymb,amsthm,mathtools,bm}
\usepackage{booktabs,longtable,tabularx,array}
\usepackage{float, authblk}
\usepackage{graphicx,subcaption}
\usepackage[dvipsnames]{xcolor}
\usepackage[round,authoryear]{natbib}
\usepackage{enumitem}
\usepackage{url}
\usepackage[colorlinks=true,allcolors=MidnightBlue]{hyperref}
\usepackage[nameinlink,noabbrev]{cleveref}
\usepackage{caption, algorithm,algpseudocode}

\graphicspath{{figures/}}
\setlist{nosep,leftmargin=*}
\newcolumntype{Y}{>{\raggedright\arraybackslash}X}

\newtheorem{theorem}{Theorem}[section]
\newtheorem{proposition}[theorem]{Proposition}

\newtheorem{lemma}[theorem]{Lemma}
\newtheorem{definition}{Definition}[section]

\newtheorem{assumption}[theorem]{Assumption}
\newtheorem{remark}[theorem]{Remark}

\newcommand{\R}{\mathbb R}
\newcommand{\N}{\mathcal N}

\newcommand{\argmin}{\operatorname*{arg\,min}}

\newcommand{\Prb}{\mathbb P}
\newcommand{\E}{\mathbb E}
\newcommand{\Var}{\text{Var}}

\title{Sketching in Regression: Sketchy in Practice?
}
\author[1]{Somarddha Das}
\author[1]{Soupayan Das Gupta}
\author[1]{Koustav Goswami}

\affil[1]{Indian Statistical Institute, Kolkata}
\affil[1]{\texttt{dassomardha2004@gmail.com, dasguptasoupayan@gmail.com, goswamikoustav8@gmail.com}}

\date{August 2026}

\begin{document}

\maketitle

\begin{abstract}
The Johnson-Lindenstrauss (JL) lemma and its extensions provide tools for solving the $L_2$ regression problem in high-dimensional settings with $n>d$, by transforming a subspace of $\R^{n}$ to a subspace of $\R^{m}$, (where $m<n$) while approximately preserving the norms of elements of the subspace. In this paper, we study the theoretical guarantees made by various JL-based subspace embeddings critically and concretely to gauge their practical applicability. We derive concrete versions of results on subspace embeddings in the literature to establish lower bounds on the reduced dimension as well as to concretely analyze the complexity of the Fast JL, OSNAP Transforms among other embeddings. These analyses suggest that guaranteeing substantial computational savings while maintaining the accuracy of the sketched solution requires data dimensions larger than those typically encountered in practice. We also see that, for practically necessary accuracy requirements or for sparsity requirements for computational gain, existing theoretical guarantees lead to substantially large reduced dimensions. This leads to very stringent data requirements for establishing guarantees of any kind. 

\vspace{0.5em}
\end{abstract}

\section{Introduction}

Linear regression is one of the oldest and most fundamental problems in statistics. Given $n$ data points $(y_i,x_i)$ with $y_i\in R,x_i\in R^d$ the task is to find $\beta$ which minimizes the sum of squares 
\[
\sum_{i=1}^n (y_i - x_i^T\beta)^2.
\]
Let $Y \in \R^n$ be the data vector for the $y$-values or responses and $X \in \R^{n \times d}$ be the data matrix with each row equal to $x_i^T$. We can reformulate the regression problem as that of finding the vector in the column space closest to $Y$, i.e, to find
\[
\hat{\beta} := \argmin_{\beta \in \R^d} \|Y-X\beta\|_2^2
\]
Throughout the paper we make the following standard assumption in the statistical theory of linear regression as well as numerical linear algebra. (See \cite{rao1973linear},\cite{golub2013matrix})
\begin{assumption}
    $X$ is of full column rank, with $n>d$.
\end{assumption}
The computational cost of calculating $\hat\beta$ explicitly has a complexity of $O(nd^2)$, which is concretely derived in later sections. A problem of interest is to reduce the computational complexity of calculating $\hat\beta$. When $n$ and $d$ are moderately sized, $\hat{\beta}$ can be calculated explicitly by standard numerical approaches such as applying QR decomposition to the normal equations (see \cite{golub2013matrix}). However the complexity becomes too costly with large values of $n$ and $d$, as is seen in modern datasets. 

To tackle this, first-order iterative optimization procedures are used instead of the direct factorization approach for large datasets. Gradient Descent is a standard iterative procedure for optimization problems. However, gradient descent suffers from slower convergence rates, particularly when the matrix is ill-conditioned or has multicollinearity.

For large, sparse matrices, where matrix-vector multiplications are efficient, Krylov subspace methods are seen to be more efficient. One approach is applying the Conjugate Gradient(CG) method to the normal equations. The Least Squares with QR-factorization (LSQR) method (\cite{paige1982lsqr}), based on the Golub-Kahan bidiagonalization process, improves on the CG method and guarantees a monotonic reduction of the residual $\|Y - X\hat{\beta}_n\|$ with each iteration, making it robust. Building on this, the LSMR algorithm (\cite{fong2011lsmr}) was introduced to monotonically minimize the norm of the gradient $\|X^T(Y - X\hat{\beta}_n)\|$ rather than the residual. This property makes LSMR advantageous for highly ill-conditioned problems, as the monotonic gradient reduction gives more stable early termination. The reader is encouraged to see \cite{golub2013matrix} for further details.

For large datasets, however, all of these methods have a necessary $O(nd)$ complexity in each iteration, required for matrix-vector multiplications, which can become prohibitive with increasing number of iterations to reach a desired tolerance level. This motivates the use of randomized dimensionality reduction techniques, to reduce the complexity of operations.

The original paper by \cite{JohnsonLindenstrauss1984} gives one of the most fundamental techniques of dimension reduction while approximately preserving distances. We state it in the form given in \cite{dasgupta2003elementary}.

 \begin{theorem}[Johnson-Lindenstrauss Theorem]
For any $0 < \epsilon < 1$ and any integer $n$, let $k$ be a positive integer such that
\begin{equation}
k \ge 4(\epsilon^2/2 - \epsilon^3/3)^{-1}\ln n.
\end{equation}
Then for any set $V$ of $n$ points in $\mathbf{R}^d$, there is a map $f : \mathbf{R}^d \to \mathbf{R}^k$ such that for all $u, v \in V$,
\[
(1 - \epsilon)\|u - v\|^2 \le \|f(u) - f(v)\|^2 \le (1 + \epsilon)\|u - v\|^2.
\]
Furthermore, this map can be found in randomized polynomial time.
\label{prop:JL statement}
\end{theorem}
A comprehensive survey of the Johnson--Lindenstrauss Lemma, its variants, and its applications can be found in \cite{surveyjl}. In this paper, we mainly consider its application in linear regression.

Subspace Embedding is an extension of this idea where we approximately preserve distances between points in an entire subspace and not just a finite set. The main idea is to construct a random linear map $\Pi\in R^{m\times n}$. The maps thus constructed without the knowledge of the subspace they preserve are called oblivious subspace embeddings.

\begin{definition}
    A random matrix $\Pi\in R^{m\times n}$ is called a $(\epsilon,\delta,d)$ oblivious subspace embedding (OSE) if for any $d$ dimensional subspace $T \subseteq R^n$ it holds that

    \[
\Prb( \forall x\in T, (1-\epsilon)\|x\|\leq\|\Pi x\|\leq(1+\epsilon)\|x\|)\geq 1-\delta
\]
\end{definition}

Significant work has been done to speed up the computation of the linear regression problem using subspace embeddings. In order to speed up computation we need to multiply by $\Pi$ fast and a popular way to do that is to take $\Pi$ to be a sparse matrix. We denote by $s$ the number of non-zero entries in each column of the matrix $\Pi$. Fast JL Transforms were first introduced in 2006 by \cite{ailon2009fast} and later, \cite{sarlos} gave an algorithm using it to speed up the computation of the linear regression problems. Later, \cite{nelson2013osnap} and \cite{clarkson2017lowrank} introduced various kinds of sparse embeddings to further accelerate the computation developing algorithms to solve linear regression using matrices of sparsity 1.  Another question of interest is how small the embedding dimension $m$ can be made. It is interesting to note that the dimension requirement of the classical JL lemma is optimal upto some constants as shown by \cite{larsen2017optimality}. For oblivious subspace embeddings, \cite{Nelson14} show that if $\epsilon,\delta<1/3$ then $m=\Omega(min(n,\frac{d}{\epsilon^2}))$. Furthermore, \cite{li2022lower} has shown that if $s=1$ then $m=\Omega(\frac{d^2}{\epsilon^2\delta})$ . \cite{chenakkod_et_al:LIPIcs.ICALP.2025.55} showed an embedding attaining optimal dimension with $s=\tilde{O}(1/\epsilon)$ non zero entries per column.

\subsection{Notation}
  For the regression problem, we consider $X$ as our $n\times d$ design matrix, where $n$ is number of data points and $d$ is number of covariates (including the intercept).
  
  For all embeddings, we consider $m$ to be the reduced dimension, that is the reduced design matrix is of size $m \times d$.

  The absolute error in preservation of the norm of the vectors by an embedding is denoted by $\epsilon \in (0,1)$. 

  $\delta \in (0,1)$ represents an upper bound on the failure probability of an embedding in preserving the norms of a finite set of vectors or of a subspace. If the algorithm is run many times $\delta$ bounds the proportion of incorrect outputs. Taking $\delta$ to be very small (say, $2^{-20}$) ensures that an error practically never occurs. Further, for brevity of expression, we write Ordinary Least Squares as OLS.

  All other notation is defined as needed in subsequent sections.
  
\subsection{Outline of the Paper}

  Almost all of the main results in the literature, use asymptotic notations that hide the underlying constants, which is not always sufficient to justify the theoretical guarantees in practical data dimensions and accuracy requirements. We derive concrete versions of results on subspace embeddings, primarily using the proof techniques of the original papers in Section \ref{sec:2-Concrete Theorems for Subspace Embeddings}. We then compare the cost of computing $\hat\beta$ by calculating $(X^TX)^{-1}X^TY$ and by computing $\hat{\beta}$ for a reduced dataset after applying subspace embedding and then investigate the conditions under which applying subspace embedding reduces the computational cost of solving the Regression problem in Section \ref{sec:3-Regression using subspace embedding}.

  We further conduct these analyses for Fast JL Transforms, Embedding using a sparse matrix and then using OLS regression on transformed matrices and embeddings using sparsity 1 matrices alongwith Fast JL Transforms in Section \ref{sec:3-Regression using subspace embedding}. We also compare the the cost of finding the OLS regression solution by the LSMR algorithm on the original and reduced dataset in Section \ref{sec:4-LSMR} to investigate the conditions when subspace embeddings provide a computational boost.

  We also study the statistical properties of the estimators of $\beta$ formed by subspace embeddings in Section \ref{sec:5-Statistical Properties} and compare the relative efficiencies of the sketched estimator to the OLS estimator for different embeddings. 
\section{Concrete Theorems for Subspace Embeddings}\label{sec:2-Concrete Theorems for Subspace Embeddings}

\subsection{Bound on Reduced Dimension for IID Sketch}

Here, we state an extension of the IID sketch used in the classical JL lemma for the preservation of the norm of finitely many vectors in $\R^n$ (\cite{dasgupta2003elementary}) to embed a subspace $\mathcal{U}$ in a lower dimension while preserving the norms of all vectors in $\mathcal{U}$. Once again, we present a concrete version of the lower bound on the reduced dimension required to guarantee preservation. The proof is based on the proof given in the survey in Numerical Linear Algebra by Woodruff(\cite{woodruff2014sketching}).

For this, we first define a class of distributions $\mathcal{J}$ from which we generate the IID sketching matrix based on the following property.

\begin{definition}[$\epsilon-\delta$ JL Property]
let $G$ be a $m \times d$ random matrix with each entry an independent random variable with a common distribution in $\mathcal{D}$, $G_{ij} \sim \mathcal{D}$. Let $\Pi = \frac{1}{\sqrt{m}}G$.
$\mathcal{D}$ satisfies the $\epsilon-\delta$-JL Property if for $m \geq \frac{8}{\epsilon^2 }\log(\frac{2}{\delta})$, we have
\[
(1 - \epsilon) \| x \|_2^2 \leq \| \Pi x \|_2^2 \leq (1 + \epsilon) \| x \|_2^2
\]
with probability atleast $(1 - \delta)$, for any vector $x \in \R^d$.
\end{definition}

Define $\mathcal{J}$ to be the class of all distributions satisfying the $\epsilon-\delta$ JL property for all $\epsilon, \delta>0$. This set $\mathcal{J}$ is non-empty because taking $\mathcal{D}=\mathcal{N}(0,1)$ satisfies the above property. Before stating the theorem, we prove a preliminary lemma which will be useful in the proof.

\begin{lemma}
Let $\mathcal{U}$ be a subspace of $\R^d$, $S_{\mathcal{U}}:= \{v \in \mathcal{U}: \|v\|=1\}$ be the set of unit vectors and $\mathcal{N}$ be a $k$-net of $\mathcal{S}_{\mathcal{U}}$, for some $k>0$. Then, for any $v \in \mathcal{S}_{\mathcal{U}}$, $\exists$ a sequence $x_i \in \mathcal{N}$ and $c_i \in \R$, such that
\[
v = \sum_{j=0}^\infty c_jx_j,
\]
with $|c_j| \leq k^{j} \quad \forall j$.
\label{lem: Property of Net}
\end{lemma}

\begin{proof}
We proceed by induction. We hypothesise that for all $n$, we find constants $c_1,\dots,c_n$ such that $|c_i| \leq k^i$ and $\|v - x_0 - c_1x_1 - \dots - c_nx_n\| \le k^{n+1}$.

Fix some $v \in \mathcal{S}_{\mathcal{U}}$, then suppose we obtain $x_1,x_2, \dots , x_n \in \mathcal{N}$ such that
\[
\| v - x_0 - c_1x_1 - \dots - c_nx_n\| \leq k^{n+1}
\]
Define $r_n:=  v - x_0 - c_1x_1 - \dots - c_nx_n$, then $\exists x_{n+1} \in \mathcal{N}$ such that 
\[
\left\| \frac{r_n}{\|r_n\|} - x_{n+1} \right\| \le k
\]
and thus,
\[
\|r_n - \|r_n\| x_{n+1}\|
\le k \|r_n\|
\le k^{n+2}.
\]
Setting $c_{n+1} = \|r_n\|$, which completes the proof.
\end{proof}

Now, we state the theorem.
\begin{theorem}\label{JL subspace embedding}
Let $\mathcal{U} \subset \mathbb{R}^{n}$ be a $d$-dimensional linear subspace in $\mathbb{R}^{n}$ where $d\leq n$. If $\Pi \in \mathbb{R}^{m \times n}$ is chosen from any distribution in $\mathcal{J}$ then with probability $1 - \delta$,
\[
(1 - \epsilon) \| v \|_{2} \leq \| \Pi v \|_{2} \leq (1 + \epsilon) \| v \|_{2}
\]
for all $v \in \mathcal{U}$, as long as $m \ge \frac{648}{25\epsilon^2}\left(d\ln \left(\frac{7}{\epsilon}\right)+ \ln\left(\frac{2}{\delta}\right)\right).$
\label{thm: IID sketching bound}
\end{theorem}

\begin{proof}
Since $\Pi$ is a linear map, we first reduce the event to all unit vectors in $\mathcal{U}$, i.e, $\mathcal{S}_{\mathcal{U}} := \{v \in \mathcal{U}: \|v\| = 1\}$. Our event becomes
\[
\left\{1-\epsilon \leq \|\Pi v\| \leq 1+\epsilon \quad \forall v \in \mathcal{U} \text{ s.t. } \|v\| = 1\|\right\}
\]

Since $\mathcal{U}$ is finite dimensional, $\mathcal{S}_{\mathcal{U}}$ is compact and hence totally bounded and thus admits a $k$-net, for any $k>0$. Let us call this net $\mathcal{N}$. 

To bound the size of $\mathcal{N}$, we provide a greedy construction of it, i.e, repeatedly add a point in $\mathcal{N}$ from $\mathcal{S}_{\mathcal{U}}$ which lies a distance of $>k$ away from each of the points already in the current version. This is bound to terminate in finitely many steps due to total boundedness and yields a set $\mathcal{N} = \{x_1, \dots, x_n\}$ where any two points are atleast $k$ distance apart and it is a net by construction. 

Now consider balls $B(x_i,\frac{k}{2}), i = 1(1)n$. These are disjoint for $i\neq j$, since otherwise the points would not have been chosen by the procedure. Thus, the total volume of these balls is
\[
|\mathcal{N}|.\frac{\pi^{d/2}}{\Gamma(\frac{d}{2}+1)}.\big(\frac{k}{2}\big)^d.
\]
Further, these balls are contained in a ball in $\R^d$ of radius $1+\frac{k}{2}$. Thus
\begin{align*}
 &|\mathcal{N}|.\frac{\pi^{d/2}}{\Gamma(\frac{d}{2}+1)}.\big(\frac{k}{2}\big)^d  \leq \frac{\pi^{d/2}}{\Gamma(\frac{d}{2}+1)}(1+\frac{k}{2})^d \\
\implies & |\mathcal{N}|  \leq \big( \frac{2+k}{k}\big )^d 
\end{align*}
We now apply the JL lemma to the net $\mathcal{N}$. Using \ref{lem: Property of Net}, we first show that preserving the norm over $\mathcal{N}$ guarantees preservation of the norm over $\mathcal{S}_{\mathcal{U}}$. If we have that $1-k \leq \|\Pi x\| \leq 1 + k$ for any $x \in \mathcal{N}$. For any $v \in \mathcal{S}_{\mathcal{U}}$ and $k<1$,
\begin{align*}
\|\Pi v\| & = \|\Pi x_0 + \sum_{j=1}^\infty c_j\Pi x_j\| \\
& \leq \|\Pi x_0\| + \sum_{j=1}^\infty |c_j|\|\Pi x_j\| \\
& \leq (1+k)\left(\|x_0\| + \sum_{j=1}^\infty |c_j|\|x_j\|\right)\\
& \leq (1+k)\left(1 + \frac{k}{1-k}\right) = 1 +  \frac{2k}{1-k}
\end{align*}
And
\begin{align*}
\|\Pi v \| & \geq \|\Pi x_0\| - \sum_{j=1}^\infty |c_j|\|\Pi x_j\| \\
& \geq (1-k) - \frac{k(1+k)}{1-k} \\
& = \frac{1-3k}{1-k}\\
& = 1 - \frac{2k}{1-k}
\end{align*}

Take $k = \frac{\epsilon}{3}$. Since $\epsilon \in (0,1)$, $k <\frac{1}{3}$, thus, $\frac{2k}{1-k} \leq 3k$. Thus, choosing the reduced dimension $m$ so that  
\[
\mathbb{P}(1-k \leq \|\Pi x\| \leq 1+k, \forall x \in \mathcal{N}) \geq 1-\delta,
\]
we have 
\[
\mathbb{P}(1-\epsilon \leq \|\Pi v\| \leq 1+ \epsilon, \forall v \in \mathcal{S}_{\mathcal{U}}) \geq 1-\delta.
\]
Further, for $k<\frac{\epsilon}{3}$, $\frac{2+k}{k}<\frac{7}{3k}$. By the $\epsilon-\delta$ JL-property, we have that for $k' = 2k - k^2$, $(1-k')\|x\|^2 \le \|\Pi x\|^2 \le (1+k')\|x\|^2$ with probability $\ge 1-\delta'$, for $m \ge \frac{8}{k'^2}\log \left(\frac{2}{\delta'}\right)$. Taking $\delta' = \frac{\delta}{|\mathcal{N}|}$, we can guarantee the preservation of norm for all vectors in $\mathcal{N}$ with probaility atleast $1-\delta$, by the union bound. 

Using the upper bound on the cardinality of $|\mathcal{N}|$ and the lower bound on $m$ by the union bound, it is sufficient to for $m$ be such that:
\begin{align*}
m & \geq \frac{8}{k'^2} \left(\ln \left(\frac{7}{3k}\right)^d+ \ln\left(\frac{2}{\delta}\right)\right) \\
& = \frac{8}{(\frac{2\epsilon}{3} - \frac{\epsilon^2}{9})^2} \left(\ln \left(\frac{7}{\epsilon}\right)^d+ \ln\left(\frac{2}{\delta}\right)\right) 
\end{align*}

 Thus, it is enough to have, $m \geq \frac{8}{(\frac{5\epsilon}{9})^2} \left(d\ln \left(\frac{7}{\epsilon}\right)+ \ln\left(\frac{2}{\delta}\right)\right) = \frac{648}{25\epsilon^2}\left(d\ln \left(\frac{7}{\epsilon}\right)+ \ln\left(\frac{2}{\delta}\right)\right)$

\end{proof}

\begin{remark}\label{rem:Gaussian IID Sketch}
The covering-net argument above provides a sufficient dimension bound for an oblivious subspace embedding, but its constants may be conservative. For Gaussian sketches, sharper bounds can be obtained by directly controlling the extreme singular values of the sketch restricted to the column space  (see \cite{vershynin2018high, davidson2001local, gordon1988milman}). Specifically, see Theorem 4.6.1 of \cite{vershynin2018high}, which provides upper and lower bounds to the spectrum of $SU$ ($S$ being the sketching matrix and $U$ is a matrix with orthonormal columns) holding with high probability.
\end{remark}
 \cite{Nelson14} prove the optimality of the $\frac{d}{\epsilon^2}$ term upto a constant. $\delta$ represents the probability of the sketch failing to preserve the norms within the subspace. In the context of regression, it can be interpreted as an upper bound on the probability that applying the sketching transform fails to yield an accurate solution.

\subsection{Concrete Lower Bound on the reduced dimension}

The paper \cite{Nelson14} shows that any OSE with $\delta<1/3$ we must have $m=\Omega(min(n,\frac{d+\log(\frac{1}{\delta})}{\epsilon^2}))$. The proof initially shows $m=\Omega(min(n,\log(\frac{d}{\delta}){\epsilon^{-2}}))$. We shall explicitly derive the second bound with all constants. The proof is done by calculating the constants at each step from \cite{Nelson14} and \cite{kane2014almost}

\begin{theorem}\label{thm:Concrete reduced dimension bounds}
     If $S:R^n \rightarrow R^m $ is a linear transformation with $n>2m$ and $\epsilon>0$. Let $\omega$ be a randomly chosen vector in $\mathcal S^{n-1} $. Additionally assume $\epsilon+\frac{1}{\sqrt{m}}<1/2$ and $m\geq 4$, then
    \[\Prb[ |\left\lVert Sw \right\rVert^2 -1|> \epsilon]\geq\frac{e^{-\frac{5}{2}(m\epsilon^2+1)}\sqrt{\pi}}{2e^{\frac{49}{12}}}.\]
\end{theorem}

\begin{proof}
 Let S be surjective without loss of generality and define $V=ker(S)$ and $U=V^{\perp}$.\\
Then $\R^n$ can be written as direct sum of $V$ and $U$. Therefore, $w \in \mathcal S^{n-1}$ can be written uniquely as $\sqrt{s_V}\Omega_V+\sqrt{s_U}\Omega_U$ where $s_U, s_V \in [0,1] \text{ such that } s_U+s_V=1$ and $\Omega_V, \Omega_U$ are unit vectors in corresponding subspaces.\\
Let $\sigma$, $\sigma_U$, and $\sigma_V$ denote the normalized surface measures on
$\mathcal S^{n-1}$, $S(U)$, and $S(V)$, respectively. By the standard decomposition of the
uniform measure on the sphere, $d\sigma(w)=f(s_U)\,ds_U\,d\sigma_U(\Omega_U)\,d\sigma_V(\Omega_V)$,where $f(s)=\frac{s^{m/2-1}(1-s)^{(n-m)/2-1}}
{B\!\left(\frac{m}{2},\frac{n-m}{2}\right)}.
$
We alternatively write $f(s)=\frac{s^{a-1}(1-s)^{b-1}}{B(a,b)},
a=\frac{m}{2},
b=\frac{n-m}{2}.$
Assume \(n>2m\), so that \(a\le b\). The density is maximized at $ s_0=\frac{m-2}{n-4}=\frac{a-1}{a+b-2}.$
Define $g(x)=\log\bigl(f(s_0(1+x))\bigr).$
Then$$
\begin{aligned}
g(x)
&=(a-1)\log\!\bigl(s_0(1+x)\bigr)
 +(b-1)\log\!\bigl(1-s_0(1+x)\bigr)
 -\log B(a,b) \\
\end{aligned}
$$
Differentiating we get, $ g'(0)
=(a-1)-(b-1)\frac{s_0}{1-s_0}
=(a-1)-(b-1)\frac{a-1}{b-1}
=0,
$
since $\frac{s_0}{1-s_0}=\frac{a-1}{b-1}.$

    Note $f(s)$ attains its maximum value at $s_0=\frac{m-2}{n-4}<1/2.$ For \(0<x\le \tfrac12\), Taylor's theorem gives
\[
g(-x)-g(0)=\frac12 g''(\xi)x^2,
\qquad
\xi\in[-x,0]\subseteq\left[-\frac12,0\right].
\]
Since \(\xi\ge -\tfrac12\),$1+\xi\ge \frac12,
\text{so}\
\frac{1}{(1+\xi)^2}\le 4.$
Also, because \(\xi\le 0\),$1-s_0(1+\xi)\ge 1-s_0,$ and hence $\frac{1}{\left(1-s_0(1+\xi)\right)^2}
\le
\frac{1}{(1-s_0)^2}.
$
Therefore,
\[
\begin{aligned}
g''(\xi)
&=
-\frac{a-1}{(1+\xi)^2}
-(b-1)\frac{s_0^2}{\left(1-s_0(1+\xi)\right)^2} \ge
-4(a-1)
-\frac{(a-1)^2}{b-1},
\end{aligned}
\]
Since \(a\le b\), $\frac{a-1}{b-1}\le 1$, and therefore
\[
g''(\xi)\ge -5(a-1)
\ge -5\left(\frac{m}{2}-1\right)
\ge -\frac{5m}{2}.
\]
Consequently, $g(-x)-g(0)
\ge
-\frac{5}{4}mx^2$,
Hence we get,
$f\bigl(s_0(1-x)\bigr)
\ge
f(s_0)\exp\!\left(-\frac54mx^2\right).$

Assume \(a,b>1\). The mode of the Beta density is $s_0=\frac{a-1}{a+b-2}.$ $f(s_0)=\frac{\Gamma(a+b)}{\Gamma(a)\Gamma(b)}
\cdot
\frac{(a-1)^{a-1}(b-1)^{b-1}}
     {(a+b-2)^{a+b-2}}.$

For every \(x\ge1\), Stirling's bounds give
\[
\sqrt{2\pi}\,x^{\,x-\frac12}e^{-x}
\le
\Gamma(x)
\le
e\,x^{\,x-\frac12}e^{-x+\frac{1}{12x}}.
\]

Hence
\[
\frac{\Gamma(a+b)}{\Gamma(a)\Gamma(b)}
\ge
\frac{\sqrt{2\pi}}{e^2}
\cdot
\frac{(a+b)^{a+b-\frac12}}
     {a^{a-\frac12}b^{b-\frac12}}
\,
e^{-\frac1{12a}-\frac1{12b}}.
\]

Substituting into the expression for \(f(s_0)\),
\[
f(s_0)
\ge
\frac{\sqrt{2\pi}}{e^2}
\cdot
\frac{(a+b)^{a+b-\frac12}}
     {a^{a-\frac12}b^{b-\frac12}}
\cdot
\frac{(a-1)^{a-1}(b-1)^{b-1}}
     {(a+b-2)^{a+b-2}}
\,
e^{-\frac1{12a}-\frac1{12b}}.
\]

 \[
f(s_0) \geq {\frac{\sqrt{2\pi}}{e^2}} \sqrt{\frac{(a+b)^3}{ab}} {\frac{\left(1-\frac{1}{a}\right)^{a-1} \left(1 - \frac{1}{b}\right)^{b-1}}{\left(1 - \frac{2}{a+b}\right)^{a+b-2}}}e^{-1/12}
\]

$\left( 1 - \frac{1}{x} \right)^{x-1} \geq e^{-1}$ $ \forall x\geq 1$ and $\left( 1 - \frac{2}{x} \right)^{(x-2)} \leq 1$ $\forall x\geq 2.$ Substituting these bounds we get

\[
f(s_0)\geq {\frac{\sqrt{2\pi}}{e^{\frac{49}{12}}}} \sqrt{\frac{(a+b)^3}{ab}} \geq {\frac{\sqrt{2\pi}}{e^\frac{49}{12}}} \sqrt{\frac{n^3}{2 m(n-m)}} \geq {\frac{\sqrt{\pi}}{e^\frac{49}{12}}} \sqrt{\frac{n^3}{ m(n-m)}}
\]

We consider the probability $\Prb(s < (1 - \epsilon) s_0)$. Expressing this as an integral over the density function $f(s)$, we have:
\[
\Prb(s < (1 - \epsilon) s_0) = \int_{0}^{(1 - \epsilon) s_0} f(s) \, ds= s_0 \int_{\epsilon}^{1} f(s_0 (1 - x)) \, dx \geqslant s_0 \int_{\epsilon}^{1/2} f(s_0 (1 - x)) \, dx
\]

Using the bound $f(s_0(1 - x)) \geqslant f(s_0) e^{-\frac{5}{4} m x^2}$, we get
\[
\Prb(s < (1 - \epsilon) s_0) \geqslant s_0 \int_{\epsilon}^{1/2} f(s_0) e^{-\frac{5}{4} m x^2} \, dx = s_0 f(s_0) \int_{\epsilon}^{1/2} e^{-\frac{5}{4} m x^2} \, dx
\]

Assume $\epsilon + \frac{1}{\sqrt{m}} < \frac{1}{2}$, we can further shrink the domain of integration to $[\epsilon, \epsilon + \frac{1}{\sqrt{m}}]$:
\[
\Prb(s < (1 - \epsilon) s_0) \geqslant s_0 f(s_0) \int_{\epsilon}^{\epsilon + \frac{1}{\sqrt{m}}} e^{-\frac{5}{4} m x^2} \,\geqslant s_0f(s_0)\frac{1}{\sqrt{m}} e^{-\frac{5}{4} m \left(\epsilon + \frac{1}{\sqrt{m}}\right)^2} \geqslant s_0f(s_0)\frac{1}{\sqrt{m}}e^{-\frac{5}{2} (m \epsilon^2 + 1)}dx
\]

Since $s_0 = \frac{m - 2}{n - 4}$ ,$s_0 f(s_0) \geqslant \frac{\sqrt{\pi}}{e^{49/12}} \cdot \frac{m - 2}{n - 4} \cdot \sqrt{\frac{n^3}{m(n - m)}}= \frac{ \sqrt{\pi}}{e^{49/12}} \cdot \frac{m - 2}{\sqrt{m}} \cdot \frac{n^{3/2}}{(n - 4) \sqrt{n - m}}\geqslant \frac{ \sqrt{\pi}}{e^{49/12}} \cdot \frac{m - 2}{\sqrt{m}}$\\
Since $m - 2 \geqslant \frac{m}{2}$, substituting this gives $s_0 f(s_0) \geqslant \frac{ \sqrt{\pi}}{2 e^{49/12}} \sqrt{m}.$
Similarly, the same bound holds for
$\Prb(s>(1+\epsilon)s_0)$.

Now, let $c=1/\|S\Omega_U\|^2$. Then
\[
\Prb\left(
|\|Sw\|^2-1|>\epsilon
\mid \Omega_U,\Omega_V
\right)
=
\Prb\left(
\|Sw\|^2 \notin
[c(1-\epsilon),c(1+\epsilon)]
\right).
\]

Since both tails satisfy the same bound, integrating gives
\[
\Prb[ |\left\lVert Sw \right\rVert^2 -1|> \epsilon] \geq \frac{s_0 f(s_0)}{\sqrt{m}} e^{-\frac{5}{2} (m \epsilon^2 + 1) }\geq \frac{\sqrt{\pi}}{2 e^{49/12}}e^{-\frac{5}{2} (m \epsilon^2 + 1)}
\] 
\end{proof}

\begin{theorem}\label{Lower Bound for t vectors}
    \textit{Suppose $\mathcal{D}$ is a distribution over $\mathbb{R}^{m \times n}$ with the property that for any $t$ vectors $x_1, \dots, x_t \in \mathbb{R}^n$,}
\[
\underset{\Pi \sim \mathcal{D}}{\mathbb{P}} \left( \forall i \in [t], \, (1 - \epsilon)\|x_i\| \le \|\Pi x_i\| \le (1 + \epsilon)\|x_i\| \right) \ge 1 - \delta. 
\] where $\delta<1/2.$
\textit{Additionally assume $2\epsilon+\epsilon^2+\frac{1}{\sqrt{m}}<1/2$ and $m\geq 4$ then $m \geq \min \left\{ n/2, \frac{\frac{2}{5}\log\left(\frac{\sqrt{\pi}t}{4e^{49/12}\delta}\right)-1}{(2\epsilon+\epsilon^2)^2}\right\}$.}
\end{theorem}
\smallskip
\begin{proof}
     The proof uses Yao's minimax principle. That is, let $\mathcal{U}$ be an arbitrary distribution over $t$-tuples of vectors in $\mathcal S^{n-1}$. Then
\[
\underset{(x_1, \dots, x_t) \sim \mathcal{U}}{\mathbb{P}} \,\, \underset{\Pi \sim \mathcal{D}}{\mathbb{P}} \left( \forall i \in [t], \, | \|\Pi x_i\|^2 - 1 | \le 2\epsilon+\epsilon^2 \right) \ge 1 - \delta.
\]
Switching the order of probabilistic quantifiers, an averaging argument implies the existence of a fixed matrix $\Pi_0 \in \mathbb{R}^{m \times n}$ so that
\[
\underset{(x_1, \dots, x_t) \sim \mathcal{U}}{\mathbb{P}} \left( \forall i \in [t], \, | \|\Pi_0 x_i\|^2 - 1 | \le 2\epsilon+\epsilon^2 \right) \ge 1 - \delta.
\]
For the condition in the theorem to hold, the above statement must be true for any distribution $\mathcal{U}$ over $\R^t$, we consider the case when $\mathcal{U}$ is the distribution of $t$ independent copies of the distribution used in \ref{thm:Concrete reduced dimension bounds}. By the previous theorem $\Pi_0$ can satisfy this equation if 
\[1-\delta\leq (1-\frac{e^{-\frac{5}{2}(m(2\epsilon+\epsilon^2)^2+1)}\sqrt{\pi}}{2e^{49/12}})^T\]

Let $(\frac{e^{-\frac{5}{2}(m(2\epsilon+\epsilon^2)^2+1)}\sqrt{\pi}}{2e^{49/12}})=\delta'$. Hence we get $1-\delta \leq (1-\delta')^t\leq e^{-t\delta'}$. So finally we get $\delta'\leq -\frac{\log(1-\delta)}{t}$.
Substituting the values we get 
\[m\geq \frac{\frac{2}{5}\log\left(\frac{\sqrt{\pi}t}{4e^{49/12}\delta}\right)-1}{(2\epsilon+\epsilon^2)^2}.\]
\end{proof}

\begin{theorem}
     Any OSE distribution $\mathcal{D}$ over $\mathbb{R}^{m \times n}$ preserving a $d$ dimensional subspace such that  $\epsilon+\frac{1}{\sqrt{m}}<1/2$ and $m\geq 4$ , we must have $m\geq \min \left\{ n/2, \frac{\frac{2}{5}\log\left(\frac{\sqrt{\pi}d}{4e^{49/12}\delta}\right)-1}{(2\epsilon+\epsilon^2)^2}\right\} $.
\end{theorem}

\begin{proof}

 We have that for any $d$-dimensional subspace $W \subset \mathbb{R}^n$, a random $\Pi \sim \mathcal{D}$ with probability $1 - \delta$ simultaneously preserves norms of all $x \in W$ up to $1 \pm \epsilon$. Thus for any set of $d$ vectors $x_1, \dots, x_d \in \mathbb{R}^n$, a random such $\Pi$ with probability $1 - \delta$ simultaneously preserves the norms of these vectors since it even preserves their span. The lower bound then follows by Theorem \ref{Lower Bound for t vectors}. 
\end{proof}
\begin{remark}
 We can make the bound stronger by modifying the assumption $2\epsilon+\epsilon^2+\frac{1}{\sqrt{m}}<1/2$ to a number much less than $1/2$ which is generally true in practice but the important dependence is the inverse proportionality to $\epsilon^2$.   
\end{remark}
 
\subsection{Lower Bound for Sparsity 1 Embeddings}
  The paper \cite{li2022lower} gives a lower bound for a sparsity 1 embedding by constructing a hard-to-preserve random subspace. 
  To show the hard instance, we use the following definition from the same paper.
  \begin{definition}[Distribution \(\mathcal{D}_\beta\)]
Let \(0<\beta\le 1\), with \(1/\beta\) an integer. The distribution \(\mathcal{D}_\beta\) on matrices \(U\in\mathbb{R}^{n\times d}\) is defined by
\[
U=VW,
\]
where \(V\in\mathbb{R}^{n\times(d/\beta)}\) and
\(W\in\mathbb{R}^{(d/\beta)\times d}\).

The columns of \(V\) are independent and identically distributed: for each
\(j\in\{1,\ldots,d/\beta\}\), the column \(V_{*,j}\) is chosen uniformly from the
\(n\) canonical basis vectors of \(\mathbb{R}^n\).

Let \(\sigma_1,\ldots,\sigma_{d/\beta}\) be independent Rademacher random
variables. For each \(i\in\{1,\ldots,d\}\), set
\[
W_{j,i}=\sigma_j\sqrt{\beta}
\qquad\text{for }\quad
j=\frac{i-1}{\beta}+1,\ldots,\frac{i}{\beta},
\]
and set all remaining entries of \(W\) to zero.
\end{definition}
Although the definition requires $1/\beta$ to be an integer, the results would still hold by taking integer part of $1/\beta$, so throughout the section we assume $1/\beta$ to be an integer, without loss of generality. For our case, the hard instance $\mathcal{D}$ is $\mathcal{D}_1$ with probability 1/2 and $\mathcal{D}_{8\epsilon}$ with probability 1/2.
We will use slightly modified versions the following lemmas given as Lemmas 6 and 7 of \cite{li2022lower}.
\begin{lemma}\label{lem:LiLiu6}
    If \(\Pi\in\mathbb{R}^{m\times n}\) is an
\((\epsilon,\delta)\)-subspace embedding for \(U\sim \mathcal{D}\) with
column sparsity \(s=1\), then at least a
\(\left(1-\frac{4\delta}{d}\right)\)-fraction of the columns of \(\Pi\)
have \(L_2\)-norm in \([1-\epsilon,1+\epsilon]\).
Equivalently, at least a \(\left(1-\frac{4\delta}{d}\right)\)-fraction
of its nonzero entries have absolute value in
\([1-\epsilon,1+\epsilon]\).
\end{lemma}
For each $i\in[m]$, $B_i$ is the number of $j \in [n]$ such that $|\Pi_{ij}|\in [1-\epsilon, 1+\epsilon]$ and $\Pi_{ij}V_{jl}\neq0$ for some $l\in [d]$. Therefore, $\Pi$ is like a hashing matrix, hashing $n$ objects into $m$ buckets and $B_i$ is the number of distinct objects hashed into the $i$-th bucket.
\begin{lemma}\label{lem:LiLiu7}
    Let \(\epsilon\in(0,1/8)\), \(\delta\in(0,1/4)\).
If \(\Pi\in\mathbb{R}^{m\times n}\) is an
\((\epsilon,\delta)\)-subspace embedding for \(U\sim \mathcal{D}\) with
column sparsity \(s=1\), then, conditioned on \(U\sim \mathcal{D}_{8\epsilon}\),
\[
\Prb\!\left(\exists\,i\text{ such that }B_i>1| U\sim \mathcal{D}_{8\epsilon}\right)
\leq 8\delta+\Prb(\mathcal{B}|U\sim \mathcal{D}_{8\epsilon})
\]
where $\mathcal{B}$ is the event that $V$ has atleast 2 identical columns.
\end{lemma}
\begin{proof}
    By Law of Total probability, we have $\Prb(\Pi \text{ is a subspace embedding for }U| U\sim \mathcal{D}_{8\epsilon})\geq 1-2\delta. $
    Let $\mathcal{E}$ be the event that $\exists\,i\text{ such that }B_i>1$.
    
    For a fixed $V\in \mathcal{E}\cap \mathcal{B}^c$, there exist 2 distinct columns $i\neq j$ of $\Pi V$ such that $|\langle(\Pi V)_{*,i}, (\Pi V)_{*,j}\rangle|\geq (1-\epsilon)^2>3/4.$ Invoking Lemma 4 of \cite{li2022lower} with $\beta=8\epsilon$ and $\lambda=6$ gives an embedding failure with probability atleast $1/4$ over randomness of $W$. Averaging over $V$, we get 
    \[ \Prb(\mathcal{E}|\mathcal{B}^c,U\sim \mathcal{D}_{8\epsilon})/4 \leq  \Prb(\Pi \text{ is not a subspace embedding for }U| U\sim \mathcal{D}_{8\epsilon})\leq 2\delta.\]
    Therefore, using the union bound on $\mathcal{E} = (\mathcal E \cap \mathcal B) \cup (\mathcal E \cap \mathcal B^c)$, we get,
    \[\Prb(\mathcal{E}|U\sim \mathcal{D}_{8\epsilon})\leq 8\delta +\Prb(\mathcal{B}|U\sim \mathcal{D}_{8\epsilon})\]
    
\end{proof}
Let $\mathcal{B}$ be the event that $V$ has 2 identical columns. Similar to Li and Liu's paper, we use the subspace property of $U$ conditioned on $\mathcal{B}^c$ and adding to their proof, we explicitly bound $\Prb(\mathcal{B}|U \sim \mathcal D_{8\epsilon})$.

For the lower bound, the hard instance $\mathcal{D}$ is $\mathcal{D}_1$ with probability $1/2$ and $\mathcal{D}_{8\epsilon}$ with probability $1/2$.
By an averaging argument, if a random matrix $\Pi$ preserves every fixed $d$-dimensional subspace with probability at least $1-\delta$, then, for any distribution over $d$-dimensional subspaces, there exists a fixed realization $\pi$ of $\Pi$  that preserves a random subspace drawn from this distribution with probability at least $1-\delta$. Li and Liu exploited this idea in their proof. We obtain our theorem by using a concrete version of the Birthday Paradox Bound in our proof.
\begin{theorem}\label{sparsity 1 lower bound}
Let \(\epsilon\in(0,1/8)\), \(\delta\in(0,1/13),d\geq 2\) and
\[
n\geq \frac{d^2}{64\epsilon^2\delta}.
\] If
\(\Pi\in\mathbb{R}^{m\times n}\) is an
\((\epsilon,\delta,d)\)-subspace embedding for \(U\sim \mathcal{D}\) with
column sparsity \(s=1\), then
\[
m\ge   \frac{d}{32\epsilon}
        \left(\frac{d}{16\epsilon}-1\right)
        \frac1{\log\left(\frac{1}{1-13\delta}\right)}
        .
\]
\end{theorem}
\begin{proof}
    From Lemma \ref{lem:LiLiu6}, at most a $\frac{4\delta}{d}$-fraction of the
    columns of $\Pi$ are bad. Conditioned on
    $U\sim\mathcal{D}_{8\epsilon}$, the expected number of selected
    bad columns is therefore at most $\frac{d}{8\epsilon}\cdot\frac{4\delta}{d}
        =\frac{\delta}{2\epsilon}.$ For the rest of the proof, we assume $U\sim\mathcal{D}_{8\epsilon}$.
    
    By Markov's inequality, with probability at least
    $1-\frac{8\delta}{d}$, at least $N:=\frac{d}{16\epsilon}$ selected columns are good, we denote this event by $A$. Let $C$ be the event that atleast $N$ good columns hash to distinct rows.
    Then $C^c \subseteq A^c\cup \mathcal{E}\cup \mathcal{B}$, where $\mathcal{E}$ is as defined in Lemma \ref{lem:LiLiu7}.
    Now, 
    Since $V$ has $r=d/(8\epsilon)$ independent columns,
    \[\Prb(\mathcal{B})\leq \frac{\binom{r}{2}}{n}\leq \frac{d^2}{128\epsilon^2n}\leq \delta/2.\]
    By Union Bound and Lemma \ref{lem:LiLiu7},
    \[\Prb(C^c) \leq \frac{8\delta}{d}+8\delta+2\Prb(\mathcal{B})\leq 13\delta\]
    By the birthday paradox,
    \[
        \Prb\left(X_i=X_j\text{ for some }i\neq j\right)
        \ge
        1-\exp\left(-\frac{N(N-1)}{2m}\right).
    \]
    This remains true for non-uniform hashing, since non-uniformity
    can only increase the collision probability. Therefore,
    \[
        1-\exp\left(-\frac{N(N-1)}{2m}\right)\le 13\delta,
    \]
    which gives
    \[
        m\ge
        \frac{
        \frac{d}{16\epsilon}
        \left(\frac{d}{16\epsilon}-1\right)
        }{
        -2\log(1-13\delta)
        }.
    \]
\end{proof}

\section{Linear Regression using Sketching} \label{sec:3-Regression using subspace embedding}

The use of subspace embeddings, more popularly known as sketching in the literature, for dimensionality reduction has led to their use in solving the linear regression problem. Various sketching algorithms and their modifications have been proposed by \cite{sarlos},\cite{sketch},\cite{nelson2013osnap},\cite{clarkson2017lowrank},\cite{chenakkod_et_al:LIPIcs.ICALP.2025.55} to obtain efficient algorithms for linear regression to make it computationally more efficient. We perform a concrete analysis of the number of computations in the worst case needed to perform linear regression by explicitly evaluating $\hat{\beta}=(X^TX)^{-1}X^TY$ and those using subspace embedding and compare them to investigate conditions under which sketching achieves computational gain . We divide the rest of the sections to examine these algorithms one by one.

\begin{itemize}
    \item Reducing the dimension and explicitly calculating $\hat{\beta}$ in the reduced dimension.

    \item Using Fast JL transforms.

    \item Using sparse matrices.

\end{itemize}

The main idea of the algorithm is using a subspace embedding to preserve the norms of vectors belonging to a specific subspace. In the context of regression, let \( E := span\{X:b\} \) denote the $d+1$-dimensional subspace spanned by the columns of \( X \) and the vector \( b \). We use a sketching matrix $\Pi$ on the $d+1$-dimensional subspace $E \subseteq \R^n$ to a $d+1$-dimensional subspace of $E' \subseteq \R^m$, for some $m<n$, by transforming,
\[ \{X^{(1)}, X^{(2)}, \dots, X^{(d)}, Y\} \quad \text{to} \quad \{X'^{(1)}, X'^{(2)}, \dots, X'^{(d)}, Y'\},\]

where $X'^{(i)} := \Pi X^{(i)}$, $i \in [d]$ and $Y'=\Pi Y$.

After projection, we solve the reduced least squares problem:

\[ \min_{\beta' \in \mathbb{R}^d} \|X'\beta' - Y'\|_2. \]
This method is also termed as the Sketch-and-Solve approach, and it is the primary object of our analysis in this paper. Another popular approach is Precondition-and-Solve where the data matrix is reduced using subspace embedding and the QR decomposition of the result is calculated. Algorithms such as Blendenpik (\cite{avron2010blendenpik}) and LSRN (\cite{meng2014lsrn}) are examples of this approach. Some other projection-based methods can be found in \cite{halko2011finding}, \cite{martinsson2020randomized}. Finally, a new regression problem is solved using  iterative methods, where the transformed data matrix has a bounded condition number.

\begin{remark}
Here reduced dimension does not mean that the dimension of the data points are being reduced .We begin with $n$ data points $x_1,x_2,...,x_n$ in $d$ dimension. We have matrix A with $i$th row $x_i^T$ . After the reduction we have $m$ data points in $d$ dimension.
\end{remark}

\begin{lemma}
With probability at least \( 1 - \delta \), the embedding $\Pi$ preserves the square root of Sum of Squares of Errors(SSE) up to a \( (1 \pm \epsilon) \) \text{factor}, that is,

\[(1 - \epsilon) \min_{\beta \in \mathbb{R}^d} \|X\beta - Y\|_2 \leq \min_{\beta' \in \mathbb{R}^d} \|X'\beta' - Y'\|_2 \leq (1 + \epsilon) \min_{\beta \in \mathbb{R}^d} \|X \beta - Y\|_2.\] 

as long as the reduced dimension $m \ge \frac{648}{25\epsilon^2}
\left[
(d+1) \log\!\left(\frac{7}{\epsilon}\right)
+ \log\!\left(\frac{2}{\delta}\right)
\right].$

\end{lemma}
    
\begin{proof}
With probability at least \( (1 - \delta) \), the subspace embedding guarantee ensures that for all vectors \( z \in E \),

\[(1 - \epsilon) \|z\|_2 \leq \|\Pi z\|_2 \leq (1 + \epsilon) \|z\|_2.\]

Fix $\pi$ to be an instance of $\Pi$ which preserves the norms upto a factor of $\epsilon$. Define

\[\beta_1^* = \arg \min_{\beta \in \mathbb{R}^d} \|X\beta - Y\|_2, \quad \beta_2^* = \arg \min_{\beta' \in \mathbb{R}^d} \|X'\beta' - Y'\|_2.\]

Let \( z = X\beta_1^* - Y \). Since \( z \in E \), it follows that

\[\|X'\beta_1^* - Y'\|_2 = \|\Pi z\|_2 \leq (1 + \epsilon) \|z\|_2 = (1 + \epsilon) \|X\beta_1^* - Y\|_2. \tag{1}\]

By optimality of \( \beta_2^* \), 
$\|X'\beta_2^* - Y'\|_2 \leq \|X'\beta_1^* - Y'\|_2\leq (1 + \epsilon) \|X\beta_1^* - Y\|_2$

Also, 

\[\| X'\beta_2^* - Y' \|_2 \geq (1 - \epsilon) \| X \beta_2^* - Y \|_2 \geq (1 - \epsilon) \| X \beta_1^* - Y \|_2\tag{2}\]

Combining the two inequalities yields the desired approximation guarantee:

\[(1 - \epsilon) \|X\beta_1^* - Y\|_2 \leq \|X'\beta_2^* - Y'\|_2 \leq (1 + \epsilon) \|X\beta_1^* - Y\|_2.\]

This completes the proof. Hence the square root of $SSE=\min_{\beta \in \mathbb{R}^d} \|X\beta - y\|_2 $ is preserved upto a factor of $\epsilon$
    
\end{proof}
\subsection{Interpretation of $\epsilon,\delta$}

Here $\delta$ represents the probability with which the designed algorithm preserves the SSE upto a factor of $(1\pm\epsilon)$ . It is the failure probability of our algorithm so we would want it to be small enough to avoid wrong instances. Checking the correctness of an output incurs a heavy cost of computation.

$\epsilon$ controls the factor by which the square root of SSE is preserved. For our algorithm to give meaningful results we would desire $\epsilon$ to be very small so that the SSE is almost preserved.

\subsection{Reducing the dimension and explicitly calculating $\hat{\beta}$ in the reduced dimension.}

We reduce the dimension by using subspace embedding and then in the lower dimension we explicitly calculate the new $\hat{\beta}$

\subsubsection{Concrete Analysis of Time complexity}
We compare the worst-case time complexities of computing the OLS estimates and the sketched estimate. Let $f(n)$ be the number of computations required to invert a $n \times n$ matrix. Let $C_{reg}$ be the run-time of canonical regression and $C_{emb}$ be the run-time of regression with subspace embedding using a dense matrix. 

\[C_{reg}= (2n-1)d^2+f(d)+(2n-1)d+(2d-1)d\]
\[C_{emb}= (2n-1)md+(2n-1)m+(2m-1)d^2+f(d)+(2m-1)d+(2d-1)d \]
Clearly we can see $C_{emb}>C_{reg}$ . Hence we can see dense matrices do not help in gaining computational efficiency. Hence we want to have a sparse embedding. Let $s$ be the number of non zero entries in each column of my embedding matrix $\Pi$.

\[C_{reg}= (2n-1)d^2+f(d)+(2n-1)d+(2d-1)d\]
\[C_{emb}= (2n-1)sd+(2n-1)s+(2m-1)d^2+f(d)+(2m-1)d+(2d-1)d \]

Note that if $s>d$ then $C_{emb}>C_{reg}$ and we do not get any computational benefit.

\cite{nelson2013osnap} introduced the OSNAP maps. Theorem 7 of \cite{chenakkod_et_al:LIPIcs.ICALP.2025.55} gives the best known result of sparsity while achieving optimal reduced dimension using the OSNAP maps. 

\begin{definition}
    (OSNAP). An $m \times n$ random matrix $S$ is called an \textit{unscaled oblivious sparse norm-approximating projection with $K$-wise independent subcolumns ($K$-wise independent unscaled OSNAP)} with parameters $p, \epsilon, \delta \in (0,1]$ such that $s = pm$ divides $m$ if,
\[
S = \sum_{l=1}^n \sum_{\gamma=1}^s \xi_{(l,\gamma)} e_{\mu_{(l,\gamma)}} e_l^T
\]
where,
\begin{itemize}
    \item $\{\xi_{(l,\gamma)}\}_{l \in [n], \gamma \in [s]}$ is a collection of $K$-wise independent Rademacher random variables.
    \item $\{\mu_{(l,\gamma)}\}_{l \in [n], \gamma \in [s]}$ is a collection of $K$-wise independent random variables such that each $\mu_{(l,\gamma)}$ is uniformly distributed in $[(m/s)(\gamma-1)+1 : (m/s)\gamma]$.
    \item The collection $\{\xi_{(l,\gamma)}\}_{l \in [n], \gamma \in [s]}$ is independent from the collection $\{\mu_{(l,\gamma)}\}_{l \in [n], \gamma \in [s]}$.
\end{itemize}
In this case, $\Pi = (1/\sqrt{pm})S$ is called a \textit{$K$-wise independent OSNAP with parameters $p, \epsilon, \delta$}. In addition, if all the random variables in the collections $\{\xi_{(l,\gamma)}\}_{l \in [n], \gamma \in [s]}$ and $\{\mu_{(l,\gamma)}\}_{l \in [n], \gamma \in [s]}$ are fully independent, then $S$ is called a fully independent unscaled OSNAP and $\Pi$ is called a fully independent OSNAP.

\vspace{1em}

Thus, each column of the OSNAP matrix $\Pi$ has $s = pm$ many non-zero entries, and the sparsity level can be varied by setting the parameter $p \in [0,1]$ appropriately. 
\end{definition}
\begin{proposition}
 (\textit{Subspace Embedding Guarantee for OSNAP}). Let $\Pi = (1/\sqrt{pm})S$ be an $m \times n$ matrix distributed according to the $8\lceil \log(\frac{d}{\delta})\rceil$-wise independent OSNAP distribution with parameter $p$. Let $U$ be an arbitrary $n \times d$ deterministic matrix such that $U^T U = I$. Then, there exist positive constants $c_{3.2.1}$ and $c_{3.2.2}$ such that for any $0 < \delta, \epsilon < 1$ and $d > 10$, we have
\[
\mathbb{P}\left(1 - \epsilon \le s_{\min}(\Pi U) \le s_{\max}(\Pi U) \le 1 + \epsilon\right) \ge 1 - \delta
\]
\textit{if the embedding dimension satisfies $m \ge c_{3.2.1}(d + \log(1/\delta\epsilon))/\epsilon^2$ and the sparsity $s = pm$ satisfies $s \ge \min\{c_{3.2.2}(\log^2(\frac{d}{\epsilon\delta})/\epsilon + \log^3(\frac{d}{\epsilon\delta})), m\}$ non-zeros per column. where $s_{min}(\Pi U)$ and $s_{max}(\Pi U)$ are the smallest and the largest singular values of $\Pi U$ .} 
\end{proposition}

\begin{remark}
The above probability is equal to the probability of subspace embedding. This is because
\begin{align*}
& \{(1-\epsilon)\|Ux\| \le \|\Pi Ux\| \le (1+\epsilon)\|Ux\| \quad \forall x\}\\
= & \{(1-\epsilon)\|x\| \le \|\Pi Ux\| \le (1+\epsilon)\|x\| \quad \forall x\}\\
= & \{1-\epsilon \le s_{min}(\Pi U) \le s_{max}(\Pi U) \le 1+\epsilon\}.
\end{align*}
The second equality follows from the orthogonality of the columns of $U$. 
\end{remark}

Note that unless the constants mentioned in the theorem are very small there is no guarantee that $s<d$ for practical datasets hence they do not guarantee faster computation.

Now we look into sparse transforms that do not attain the optimal reduced dimension. Consider the matrix $\Pi$ to have sparsity $s=1$. If we want $C_{reg}>C_{emb}$ we need
$(2n-1)d^2>(2n-1)+(2m-1)d^2+(2m-1)d$. We know from Theorem \ref{sparsity 1 lower bound} that $ m\ge
        \frac{
        \frac{d}{16\epsilon}
        \left(\frac{d}{16\epsilon}-1\right)
        }{
        -2\log(1-13\delta)
        }.$ Since $\epsilon,\delta$ have to be very small this imposes a strong lower bound on $m$. For the OSNAP construction, we have a guarantee of embedding when $m>\delta^{-1}(d^2+d)/(2\epsilon-\epsilon^2)^2$, by Theorem 3 of  \cite{nelson2013osnap}. For practical values of $d,\epsilon,\delta $ the lower bound of $m$ is large and since $n>m$ this would need a large number of data points than is generally available in practice. Hence, achieving the guarantee is not always feasible. 

\subsection{Fast JL Transform}
  Fast JL transforms were first introduced in the paper \cite{ailon2009fast}. It is a fast dimension reduction technique that combines a randomized orthogonal transformation with a sparse random projection. The main idea is to first process the data using a randomized Fourier transform that roughly distributes the mass of the vector uniformly across its coordinates. This is followed by a sparse projection that reduces the dimension while approximately preserving Euclidean distances ($L_p$ distances in particular). We will only consider $L_2$ distance in our case.

  Given $d$ points in $\R^n$ let $\Phi:\R^n \xrightarrow[]{}\R^m$ given by $\Phi= \frac{1}{\sqrt{m}}PHD$.
\begin{enumerate}
     \item $P$ is a $m \times n$ matrix with entries from a mixture of 0 and normal random variables.
     More precisely,
\[
P_{ij} \sim
\begin{cases}
\mathcal{N}(0,q^{-1}) & \text{with probability } q,\\
0 & \text{with probability } 1-q.
\end{cases}
\] and 
\[
q = \min\left\{\frac{c\log^{2} d}{n},1\right\}\quad\text{ for some $c>0$}.
\]
Here $q$ is a parameter controlling the sparsity of the matrix.
\item $H$ is the normalised $n\times n$ Hadamard matrix.
\[
H_{ij} = n^{-1/2}(-1)^{\langle i-1,j-1\rangle},
\]
where \(\langle i,j\rangle\) denotes the dot product of the \(\log n\)-bit vectors corresponding to \(i\) and \(j\) when expressed in binary.
\item $D$ is the $n\times n$ diagonal matrix, where $D_{ii}$ is $\pm 1$ each with probability $1/2$. 
 \end{enumerate}
 Both Lemma 1 of \cite{ailon2009fast} and Theorem 3 of \cite{sarlos}, state the theoretical guarantee of these transforms. We will look at the version in \cite{ailon2009fast} and prove the result with precise constants.

 An asymptotic version of the following technical lemma is stated as Lemma 2.3 in \cite{ailon2009fast}.
\begin{lemma}\label{fast lemma}
Let $u\in\mathbb R^n$ satisfy $\|u\|_2=1$ and
$\|u\|_\infty^2\le 1/k$ for some $k>0$. Let
$b_{ij}\sim\operatorname{Bernoulli}(q)$ and
$r_{ij}\sim\mathcal N(0,1/q)$, $i\in[m]$, $j\in[n]$,
be mutually independent, and define
\[
Y_i=\sum_{j=1}^n r_{ij}b_{ij}u_j,
\qquad
Z_i=\sum_{j=1}^n b_{ij}u_j^2.
\]
Then $Y_i\mid Z_i\sim\mathcal N(0,Z_i/q)$ and
$\mathbb E Z_i=q$. Suppose either $q=1$, or $0<q<1$ and
\[
q\ge \frac1k\max\left\{
\frac{32}{3}\log(80dm),\,
\frac{8}{3\epsilon^2m}\log(80d)
\right\}.
\]
Then, with probability at least $1-\frac1{20d}$,
\[
\frac q2\le Z_i\le 2q \quad(i\in[m]),
\qquad
mq(1-\epsilon)\le \sum_{i=1}^m Z_i
\le mq(1+\epsilon).
\]
\end{lemma}
\begin{proof}
Conditionally on $(b_{i1},\ldots,b_{in})$, the variable $Y_i$ is
a linear combination of independent centered Gaussian variables. Thus,
\[
Y_i\mid (b_{i1},\ldots,b_{in})
\sim \mathcal N\left(0,\frac1q\sum_{j=1}^n b_{ij}u_j^2\right)
=\mathcal N\left(0,\frac{Z_i}{q}\right).
\]
Since this conditional distribution depends on the Bernoulli variables
only through $Z_i$, we have
$Y_i\mid Z_i\sim\mathcal N(0,Z_i/q)$. Also,
$\mathbb E Z_i
=\sum_{j=1}^n \mathbb E(b_{ij})u_j^2
=q\sum_{j=1}^n u_j^2=q.$

If $q=1$, then $b_{ij}=1$ almost surely, so $Z_i=1$ for every $i$,
and both conclusions hold deterministically.

Suppose $0<q<1$, we write
\[
Z_i-q=\sum_{j=1}^n (b_{ij}-q)u_j^2.
\]
The summands are independent, centered, and satisfy
\[
\left|(b_{ij}-q)u_j^2\right|\le \frac1k,
\qquad
\sum_{j=1}^n \operatorname{Var}\bigl((b_{ij}-q)u_j^2\bigr)
=q(1-q)\sum_{j=1}^n u_j^4
\le q\|u\|_\infty^2\sum_{j=1}^n u_j^2
\le \frac qk.
\]
Bernstein's inequality therefore gives, for $0<\alpha\le1$,
\[
\Prb\left(|Z_i-q|\ge\alpha q\right)
\le 2\exp\left(-\frac38\alpha^2kq\right).
\]
Taking $\alpha=1/2$ and applying the union bound yields
\[
\Prb\left(\exists i\in[m]:Z_i\notin[q/2,2q]\right)
\le 2m e^{-\frac{3}{32}kq}.
\]

Similarly,
\[
\sum_{i=1}^m Z_i-mq
=\sum_{i=1}^m\sum_{j=1}^n (b_{ij}-q)u_j^2.
\]
The total variance is at most $mq/k$, while every summand is bounded
in absolute value by $1/k$. Hence, for $0<\epsilon\le1$,
\[
\Prb\left(
\left|\sum_{i=1}^m Z_i-mq\right|\ge\epsilon mq
\right)
\le 2\exp\left(-\frac38\epsilon^2mkq\right).
\]
Consequently, the probability that either desired conclusion fails
is at most
\[
2m e^{-\frac{3}{32}kq}
+2e^{-\frac38\epsilon^2mkq}.
\]
Under the stated lower bound on $kq$, each term is at most
$1/(40d)$, so their sum is at most $1/(20d)$,
completing the proof.
\end{proof}

Next we give a sufficient condition under which Fast JL Transform preserves distances. The result given in Lemma 1 of  \cite{ailon2009fast}, uses an asymptotic notation, so we give sufficient condition by tracking all the constants in the proof techniques.
\begin{lemma}\label{FastJl finite}
    Let $X\subseteq \R^n$ be a set of size $d$, where $2\leq n$ and $0<\epsilon<1$. Let $H,D$ be matrices as defined earlier. Let $P=(p_{ij})_{i\in[m], j\in [n]}$ where $p_{ij}=b_{ij}r_{ij}$,and $b_{ij}\sim\operatorname{Bernoulli}(q)$ and
$r_{ij}\sim\mathcal N(0,1/q)$,are mutually independent. Define 
\[\Phi= \frac{1}{\sqrt{m}}PHD\] and 
\[k=\frac{n}{2\log(40dn)}.\]
Suppose
\[
q\ge
\min\left\{
1,\,
\frac1k
\max\left[
\frac{32}{3}\log(80dm),
\frac{24}{\epsilon^2m}\log(80d)
\right]
\right\}
\]
and
\[
m\ge
\frac{216}{\epsilon^2}\log(40d),
\]
then, with probability at least \(2/3\),
\[
(1-\epsilon)\|x\|_2
\le
\|\Phi x\|_2
\le
(1+\epsilon)\|x\|_2
\qquad
\text{for every }x\in X.
\]
\end{lemma}
\begin{proof}
    The claim is obvious for $x=0$, so without loss of generality we can replace $x$ by $x/\|x\|_2$ and consider all $x$ to be of norm 1.
    For $x\in X$, let $u=HDx$.

    As $H,D$ are orthogonal, $\|u\|_2=1$.
    For every coordinate $l \in [n]$
    \[ u_l=\frac{1}{\sqrt{n}}\sum_{j=1}^n \omega_{lj}x_j,\]
    where, for fixed $l$, $\omega_{lj}$ are independent Rademacher random variables. Hence by Hoeffding's inequality,
    \[ \Prb(|u_l|\geq s) \leq 2e^{-ns^2/2}.\]
    Taking a union bound over all $x \in X$ and $l\in [n]$, and fixing 
    \[ s^2= \frac{2\log (40dn)}{n}=\frac1k,\]
    gives
    \[\Prb(\max_{x\in X} \|HDx\|_{\infty}^2\geq \frac1k)\leq 2dne^{-ns^2/2}=\frac1{20}.\]
    Let $A$ denote the event $\{\max_{x\in X} \|HDx\|_{\infty}^2< \frac1k\}$. Fix a realization of $D$ such that $A$ holds. For each $x\in X$ we define the following
    \[
Y_i=(Pu)_i
=
\sum_{j=1}^n r_{ij}b_{ij}u_j,
\qquad
Z_i=\sum_{j=1}^n b_{ij}u_j^2.
\]
By Lemma \ref{fast lemma} with $\alpha= \epsilon/3$ and the bound on $q$ from assumption we get
\[\Prb(B_x^c|A)\geq 1-\frac{1}{20d}\]
where $B_x$ denotes the event in the lemma, that is, $B_x=\{q/2\le Z_i\le2q \text{ for every }i\in[m],
mq(1-\alpha)\le\sum_{i=1}^mZ_i
\le mq(1+\alpha)\}$.
By union bound, $B_x$ holds for all $x$ with probability atleast $19/20$, and let that event be $B$.

Fix $x$ and let 
\[
v_i=\frac{Z_i}{q},
\qquad
S=\sum_{i=1}^mY_i^2,
\qquad
M=\sum_{i=1}^mv_i.
\]
Conditionally on \(Z_1,\ldots,Z_m\), the variables \(Y_i\) are
independent and $Y_i\sim\mathcal N(0,v_i)$.
On \(B\),
\[
\frac12\le v_i\le2,
\qquad
\sum_{i=1}^mv_i^2\le2M.
\]

Using the moment-generating function of a Gaussian square, for
\(0\le\lambda\le1/8\),
\[
\mathbb E\left[
e^{\lambda(S-M)}
\,\middle|\,
Z_1,\ldots,Z_m
\right]
\le
\exp\left(
2\lambda^2\sum_{i=1}^kv_i^2
\right).
\]
Taking \(\lambda=\alpha/8\) and applying Markov's inequality gives
\[
\Prb\left(S>(1+\alpha)M
\,\middle|\,
Z_1,\ldots,Z_m\right)
\le
\exp\left(-\frac{\alpha^2M}{16}\right).
\]
The analogous negative moment-generating-function argument gives
\[
\Prb\left(S<(1-\alpha)M
\,\middle|\,
Z_1,\ldots,Z_m\right)
\le
\exp\left(-\frac{\alpha^2M}{16}\right).
\]
Therefore, since \(M\ge m(1-\alpha)\),
\[
\Prb\left(
|S-M|>\alpha M
\,\middle|\,
B\right)
\le
2\exp\left(
-\frac{\alpha^2m(1-\alpha)}{16}
\right).
\]
Putting $\alpha=\epsilon/3 \leq 1/3$, we have $\alpha^2(1-\alpha)/16 \geq \epsilon^2/216$.
By the assumtion $m\ge
\frac{216}{\epsilon^2}\log(40d)$ and an union bound over $x$,
\[(1-\alpha)M \leq S \leq (1+\alpha)M \quad \text{holds for all $x\in X$ except on an event of conditional probability atmost 1/20.}\]
But on $B$, $m(1-\alpha)\leq M \leq m(1+\alpha)$.
So putting the value of $\alpha$ and replacing the square roots,
\[ (1-\epsilon) \leq \|\Phi x\|_2 \leq (1+\epsilon) \text{ for all $x$}.\]
So total failure probability is $3/20 < 1/3$ and that completes the proof.
\end{proof}
\begin{theorem}\label{FastJL thm}
Let $X\subseteq \R^n$ be a subspace of dimension $d$, where $2\leq n$ and $0<\epsilon<1$. Let $H,D$ be matrices as defined earlier. Let $P=(p_{ij})_{i\in[m], j\in [n]}$ where $p_{ij}=b_{ij}r_{ij}$,and $b_{ij}\sim\operatorname{Bernoulli}(q)$ and
$r_{ij}\sim\mathcal N(0,1/q)$,are mutually independent. Define 
\[\Phi= \frac{1}{\sqrt{m}}PHD\] and 
\[k=\frac{n}{2\log(40n)+2d\log (\frac{6+\epsilon}{\epsilon})}.\]
Suppose
\[
q\ge
\min\left\{
1,\,
\frac1k
\max\left[
\frac{32}{3}(\log(80m)+d\log (\frac{6+\epsilon}{\epsilon})),
\frac{24}{\epsilon^2m}(\log(80)+d\log (\frac{6+\epsilon}{\epsilon}))
\right]
\right\}
\]
and
\[
m\ge
\frac{216}{\epsilon^2}\left(d\log (\frac{6+\epsilon}{\epsilon})+\log 80\right),
\]
then, with probability at least \(2/3\),
\[
(1-\epsilon)\|x\|_2
\le
\|\Phi x\|_2
\le
(1+\epsilon)\|x\|_2
\qquad
\text{for every }x\in X.
\]    
\end{theorem}
\begin{proof}
    The proof follows directly from Lemma \ref{FastJl finite}, by fixing all vectors in a net of size atmost $\left(\frac{6+\epsilon}{\epsilon}\right)^d$ as in the proof of Theorem \ref{JL subspace embedding}.
\end{proof}
\begin{remark}
A related embedding called Subsampled Randomized Hadamard Transform (SRHT) is also well-known in the literature. A concrete theorem for SRHT with much better constants is given in Theorem 3.1 and 3.2 of \cite{tropp2011improved}. 
\end{remark}
\subsubsection{Concrete Time Complexity Analysis}
 In the paper \cite{sarlos}, an algorithm to solve $L_2$ regression using Fast JL Transform is introduced. Given a $n\times d$ design matrix, and response vector $y$, the algorithm simply solves the sketched regression problem of minimizing $\|\Phi X\beta- \Phi Y\|_2^2$ over $\beta$ where $\Phi$ is an $m \times n$ Fast JL matrix. The parameter $q$ is taken to be $\min\{\Theta(\frac{\log^2 N}{n}),1\}$ where $N=\max\{n,d\}$. The paper gives the theoretical guarantees and time complexity of this algorithm in Theorem 12.

 If $q=1, \Phi $ becomes a dense matrix and that reduces to an earlier case. For calculating $DX,DY$, time complexity is $n(d+1)$,  for applying the $H$ transform by Fast Fourier Transform is $n(d+1)\log_2 n$ and  the expected time complexity to multiply $P$ with the result is $qm(2n-1)(d+1)$. The expected time to generate the Gaussian and Rademacher random variables are $qmn$ and $n$ respectively. For the multiplication $\Phi X, \Phi Y$, the expected time is $qm(2n-1)(d+1)$.
 Including the time to solve $m\times d$ regression, we get,
 \[C_{fast}=n(d+1)+n(d+1)\log_2 n+qm(2n-1)(d+1)+qmn+n+(2m-1)d^2+f(d)+(2m-1)d+(2d-1)d\]

For Fast JL to be faster than usual regression, we must have $C_{reg}>C_{fast}$.
\begin{align*}
&(2n-1)d^2+f(d)+(2n-1)d+(2d-1)d\\
&>n(d+1)+n(d+1)\log_2 n+qm(2n-1)(d+1)+qmn+n\\
&+(2m-1)d^2+f(d)+(2m-1)d+(2d-1)d\\
&2(n-m)d^2+2(n-m)d>n(d+1)+n(d+1) \log_2 n+qm(2n-1)(d+1)+qmn+n
\end{align*}
Clearly if $qm>d$ then $qm(2n-1)d>2(n-m)d^2$ and also $nd\log n +nd>2(n-m)d$ So clearly RHS is greater. So for Fast JL to perform better we need $qm<d$ hence we get 
\[\frac{2\log(40n)+2d\log (\frac{6+\epsilon}{\epsilon})}{n}\cdot\frac{32}{3}(\log(80m)+d\log (\frac{6+\epsilon}{\epsilon}))m<d.\] Plugging the values of lower bounds from Theorem \ref{FastJL thm} we can observe for practical values of $d$ and small values of $\epsilon$ we need extremely large values of $n$ which is practically infeasible.

\subsection{Sparse JL Transforms}\label{sec:4-Sparse JLT}
Theorem 15 of \cite{nelson2013osnap} gives another algorithm for regression using sparsity 1 matrices. The idea is to use a sparsity 1 matrix and solve the least square problem in the reduced dimension using Fast JL as proposed in \cite{sarlos}.
\subsubsection{Concrete Time Complexity analysis}

Hence we calculate the total number of computation needed to compute $\Pi A$ and $\Pi b$ and do fast JL in the reduced dimension. Hence the total number of computation
$C_{sparse}=(2n-1)d+(2n-1)+m(d+1)+m(d+1 )\log_2 m+qm'(2m-1)(d+1)+(2m'-1)d^2+f(d)+(2m'-1)d+(2d-1)d+qmm'+m$ where $m$ is the reduced dimension after multiplying the sparsity 1 matrix and $m'$ is the further reduced dimension after Fast JL. Note that for OSNAP we require $m>\delta^{-1}(d^2+d)/(2\epsilon-\epsilon^2)^2$ by Theorem 3 of  \cite{nelson2013osnap}. In general for other sparsity 1 embeddings (like CountSketch map as given in \cite{clarkson2017lowrank}) by Theorem \ref{sparsity 1 lower bound} we have $m\ge   \frac{d}{32\epsilon}
        \left(\frac{d}{16\epsilon}-1\right)
        \frac1{\log\left(\frac{1}{1-13\delta}\right)}$ So for small values of $\epsilon$ and $\delta $, $m$ becomes extremely large. Clearly $C_{sparse}$ is much larger unless $n$ is extremely large. For $C_{reg}>C_{sparse}$ we need $(2n-1)d^2>(2n-1)+md+md \log m+qm'(2m-1)d+(2m'-1)d^2+(2m'-1)d$ for which clearly $n$ has to be huge. We can see a trivial bound that $n>m$ which in itself is a huge restriction. Hence for practical values of $n,d$, achieving any computational advantage is unlikely.

\section{Comparison with LSMR Algorithm}\label{sec:4-LSMR}

It is difficult to solve the $L_2$ regression problem exactly for large design matrices and here, iterative solvers (\cite{golub2013matrix}) or sketching can be employed in order to speed up computations. In this section, we compare the computational complexities of the best known iterative solvers --- LSQR/LSMR with the sketched complexity of these algorithms, which is, the complexity of the algorithm where we apply an iterative solver over the sketched dataset. 

The LSQR and LSMR algorithms are both iterative Krylov subspace methods based on the Golub Kahan bidiagonalization procedure, which improves on conjugate gradient approaches for large, sparse matrices (\cite{saad2003iterative}, \cite{hestenes1952methods}). The LSQR algorithm (\cite{paige1982lsqr}) is highly robust because it guarantees a monotonic reduction of the residual norm $\Vert{}Y - X\hat{\beta}_k\Vert{}$ with each iteration. Building on this framework, the LSMR algorithm (\cite{fong2011lsmr}) was introduced to monotonically minimize the norm of the gradient $\Vert{}X^T(Y - X\hat{\beta}_k)\Vert{}$ instead of the residual. This specific property makes LSMR particularly advantageous for highly ill-conditioned problems, as the monotonic gradient reduction provides a much more stable criterion for early termination. However, both algorithms have very similar time complexities. Thus, we will compare the complexity of the LSMR algorithm with the sketched LSMR algorithm.

\subsection{The LSMR algorithm and it's Complexity}

We now give a step-by-step pseudocode for the LSMR algorithm.

\begin{list}{}{\leftmargin=1em \rightmargin=1em}
\item[]
\captionsetup{type=algorithm}
\caption{Least Squares Minimum Residual (LSMR)}
\label{alg:lsmr}

\begin{algorithmic}[1]
\Require Design matrix $X \in \mathbb{R}^{n \times d}$, Response vector $Y \in \mathbb{R}^n$, Tolerance $\eta$
\State \textbf{Initialize:}
\State $\beta_1 = \|Y\|_2, \quad u_1 = Y / \beta_1$
\State $\alpha_1 = \|X^T u_1\|_2, \quad v_1 = X^T u_1 / \alpha_1$
\State $\bar{\alpha}_1 = \alpha_1, \quad \bar{\zeta}_1 = \alpha_1 \beta_1$
\State $\rho_0 = 1, \quad \bar{\rho}_0 = 1, \quad \bar{c}_0 = 1, \quad \bar{s}_0 = 0$
\State $h_1 = v_1, \quad \bar{h}_0 = \mathbf{0}, \quad x_0 = \mathbf{0}$

\For{$k = 1, 2, \dots$ until convergence $\|X^T(Y - Xx_k)\|_2 \le \eta \|X^T Y\|_2$}

    \Statex \Comment{\textit{A. Bidiagonalization}}
    \State $p = Xv_k - \alpha_k u_k$
    \State $\beta_{k+1} = \|p\|_2$
    \State $u_{k+1} = p / \beta_{k+1}$
    \State $q = X^T u_{k+1} - \beta_{k+1} v_k$
    \State $\alpha_{k+1} = \|q\|_2$
    \State $v_{k+1} = q / \alpha_{k+1}$

    \Statex \Comment{\textit{B. Orthogonalization (Givens Rotations)}}
    \State $\rho_k = \sqrt{\bar{\alpha}_k^2 + \beta_{k+1}^2}$
    \State $c_k = \bar{\alpha}_k / \rho_k, \quad s_k = \beta_{k+1} / \rho_k$
    \State $\theta_{k+1} = s_k \alpha_{k+1}$
    \State $\bar{\alpha}_{k+1} = c_k \alpha_{k+1}$

    \State $\bar{\theta}_k = \bar{s}_{k-1} \rho_k$
    \State $\bar{\rho}_k = \sqrt{(\bar{c}_{k-1} \rho_k)^2 + \theta_{k+1}^2}$
    \State $\bar{c}_k = \bar{c}_{k-1} \rho_k / \bar{\rho}_k, \quad \bar{s}_k = \theta_{k+1} / \bar{\rho}_k$
    \State $\zeta_k = \bar{c}_k \bar{\zeta}_k$
    \State $\bar{\zeta}_{k+1} = -\bar{s}_k \bar{\zeta}_k$

    \Statex \Comment{\textit{C. Search Direction and Solution Update}}
    \State $\bar{h}_k = h_k - (\bar{\theta}_k \rho_k / (\rho_{k-1} \bar{\rho}_{k-1})) \bar{h}_{k-1}$
    \State $x_k = x_{k-1} + (\zeta_k / (\rho_k \bar{\rho}_k)) \bar{h}_k$
    \State $h_{k+1} = v_{k+1} - (\theta_{k+1} / \rho_k) h_k$

\EndFor
\end{algorithmic}
\end{list}

In the above algorithm, Steps 8-13 are the Golub-Kahan Bidiagonalization process and Steps 14-22 solve a least squares subproblem, which can be solved efficiently due to bidiagonalization.

\paragraph{Computational Complexity Comparison:}
The initialization process takes $F_{init}:=2nd+3n+2d+1$ operations. Each iteration of the algorithms takes 
\[
F_{iter}:=4nd+4n+10d+24
\]
floating point operations (FLOPs). Since the LSMR algorithm is equivalent to the MINRES algorithm (\cite{paige1975solution}), i.e, it leads to the same iterate sequence, it requires at most $k \approx \frac{\kappa(X)}{2}\ln(2/\eta)$ iterations. Therefore, the global time complexity of the classical LSMR solver is:
\begin{align*}
C_{\text{LSMR}} &  \le 2nd+3n+2d+1+\left[\frac{\kappa(X)}{2}\ln\left(\frac{2}{\eta}\right)\right] \times [4nd+4n+10d+24]
\end{align*}
FLOPs, which is similar to the complexity of the LSQR algorithm, however LSMR is more robust to ill-conditioned matrices. 

Now, suppose using a sketching matrix $S$, we reduce the data matrix $X$ to $X'=SX \in \R^{m\times d}$. Applying LSMR on $X'$ upto an accuracy of $\eta$ takes about 
\[
 2md+3m+2d+1+\left[\frac{\kappa(X')}{2}\ln\left(\frac{2}{\eta}\right)\right] \times [4md + 4m + 10d+24] \text{ FLOPs},
\]
Along with this there is the added cost of sketching, which takes at most $mn(d+1)$ FLOPs. This cost can be reduced by the sparsity of the sketching matrix $S$ or the design matrix $X$. Let the complexity due to sketching be $T_{sk}$. Then total complexity of using LSMR on the sketched dataset is
\begin{align*}
C_{LSMR-JL} \le  T_{sk}+ 2md+3m+2d+1+ \left[\frac{\kappa(X')}{2}\ln\left(\frac{2}{\eta}\right)\right] \times [4md + 4m + 10d +24]
\end{align*}

We do not assume sparsity of the design matrix $X$, however  sparse sketching matrix $S$ reduces the sketching time-complexity. Say $S$ has a sparsity of $s$. Then the computation of $SX$ and $SY$ takes $2snd+2sn$ FLOPs. The Subspace Embedding guarantee ensures that for all $v$ in the subspace,
\[
(1-\epsilon)\|v\| \le \|\Pi v\| \le (1+\epsilon)\|v\|.
\]
Thus, the condition number of $X' = SX$ is  inflated as, under the subspace embedding guarantee, for all $v \in \R^d$,
\begin{align*}
(1-\epsilon)\|Xv\| \le \|X'v\| \le (1+\epsilon)\|Xv\|.
\end{align*}
Thus,
\begin{align*}
\kappa(X') & = \frac{\sigma_{max}(X')}{\sigma_{min}(X')}\\
           & \le \frac{1+\epsilon}{1-\epsilon} \frac{\sigma_{max}(X)}{\sigma_{min}(X)}\\
           & = \frac{1+\epsilon}{1-\epsilon} \kappa(X)
\end{align*}
We can hence provide an upper bound of the computational complexity of the LSMR-JL approach.
\begin{align*}
C_{LSMR-JL} & \le 2snd+2sn + 2md+3m+2d+1 + \left[\frac{\kappa(X')}{2}\ln\left(\frac{2}{\eta}\right)\right] \times [4md + 4m + 10d +24]\\
\end{align*}
We then find the regime where sketching is valid, i.e, where a computational advantage can be found: $C_{LSMR}>C_{LSMR-JL}$. Let $K = \frac{\kappa(X)}{2}\ln\left(\frac{2}{\eta}\right)$. Comparing the worst-case time complexities for both cases, we get, 
\begin{align*}
&2nd+3n+2d+1+ K \times [4nd + 4n + 10d+24] \\&> 2sn(d+1) + 2md+3m+2d+1+\frac{1+\epsilon}{1-\epsilon}K \times [4md + 4m + 10d+24] \\
\end{align*}
Simplifying, we get,
\begin{align*}
n > \frac{2md+3m + K \left\{\frac{1+\epsilon}{1-\epsilon}(4md+4m+10d+24) - 10d - 24\right\}}{2d+3+(4K-2s)(d+1)}
\end{align*}
 In the last step, we implicitly use the fact that $2d+3 + (4K-2s)(d+1) >0$, for which $s < 2K$ is a sufficient condition. This is true for sparse sketching matrices and with tolerance level $\eta = 10^{-3}$. In order to achieve computational gains, we need to apply a sparse JL transform to $X$. From Section \ref{sec:4-Sparse JLT}, we know that, for sparsity-1 JL Transforms, that $m \ge \frac{d}{32\epsilon}\left(\frac{d}{16\epsilon}-1\right)\frac1{\log\left(\frac{1}{1-13\delta}\right)}$. This imposes a steep lower bound on the number of data points $n$ for the JL-LSMR approach to yield a computational advantage. 
\section{Statistical Properties of the Sketched Estimator} \label{sec:5-Statistical Properties}
We have looked at the computational aspects of using JL Transforms in the $L_2$ regression problem, specifically focusing on whether they provide significant computational gains in minimizing the squared loss and studying the tradeoff between computational speed and approximation accuracy. However, in statistical linear regression, the primary aim is fundamentally different: computing the regression coefficient 
\[
\hat{\beta} = \arg \min_{\beta \in \mathbb{R}^d} \|Y - X\beta\|_2
\]
for a given design matrix $X \in \mathbb{R}^{n\times d}$ and response vector $Y \in \mathbb{R}^n$. 

In the classical setting, this estimator is consistent, unbiased, and, under standard regularity conditions, asymptotically normal conditioned on $X$. Furthermore, $\hat{\beta}$ is the Best Linear Unbiased Estimator (BLUE) for $\beta$ (\cite{rao1973linear, searle1971linear}), and its asymptotic normality allows for the construction of valid confidence sets. A natural question is whether these essential statistical properties hold for the sketched estimator. Given that the OLS estimator achieves the Gauss-Markov bound, we expect any sketched estimator to incur a statistical penalty. Our aim is to highlight the exact statistical tradeoffs faced when using these randomized estimators, addressing a gap in a literature that predominantly focuses on algorithmic complexity rather than inference.

\cite{Ahfock2021} provide a systematic study of the statistical properties of sketched least squares estimators. Following this, \cite{Browne2024} developed statistical inference procedures for various sketching methods, while \cite{DobribanLiu2019} and \cite{RaskuttiMahoney2016} analyzed the relative efficiency of sketched estimators. Alternative approaches to large-scale least squares include algorithmic leveraging and subsampling \cite{ma2015statistical,wang2018optimal}, bootstrap-based error estimation for sketched least squares \cite{lopes2018error}, and unbiased or debiased estimators based on volume sampling and surrogate sketching \cite{derezinski2018unbiased,derezinski2020debiasing}.

This section builds on these foundations to study the exact moments of the Gaussian sketched estimator, as well as the asymptotic distributions of the Clarkson-Woodruff and Hadamard sketched estimators, interpreting their implications for practical statistical inference.

\paragraph{Model:} We consider the classical linear model
\[
Y = X\beta_0 + \epsilon
\]
where $\epsilon \in \mathbb{R}^n$ is a random noise vector. We assume $\mathbb{E}[\epsilon] = 0$ and $\text{Var}(\epsilon) = \sigma^2I_n$ for some $\sigma>0$; essentially, the errors are zero-mean and uncorrelated. For a random sketching matrix $S \in \mathbb{R}^{m \times n}$, the sketched regression coefficient estimator is defined as:
\[
\hat{\beta}_S = (X^TS^TSX)^{-1}X^TS^TSY
\]

\subsection*{Moments of the Sketched Estimator}

We first examine the finite-sample moments of $\hat{\beta}_S$. The classical OLS estimator $\hat{\beta}$ satisfies:
\begin{enumerate}
    \item $\mathbb{E}_Y[\hat{\beta}] = \beta_0$
    \item $\text{Var}_Y[\hat{\beta}] = \sigma^2 (X^TX)^{-1}$
\end{enumerate}

To evaluate the expectation and variance of $\hat{\beta}_S$, we utilize the law of iterated expectations over the joint distribution of $(Y, S)$. 

Conditioned on $S$, the sketched estimator remains a linear function of $Y$. Therefore, taking the expectation over $Y$ first yields:
\[
\mathbb{E}_{(Y,S)}[\hat{\beta}_S] = \mathbb{E}_S\left[\mathbb{E}_Y\left[(X^TS^TSX)^{-1}X^TS^TSY \mid S\right]\right]
\]
Substituting $\mathbb{E}_Y[Y] = X\beta_0$, we see that the sketched estimator remains unbiased. We compute the variance of the sketched estimator for the Gaussian sketch, using the law of total variance and conditioning on the response $Y$. For distributional results used in the proof, refer to \cite{mardia1979multivariate}. 

\begin{theorem}
$\Var(\hat{\beta}_S) = \frac{n+m-2d-1}{m-d-1} \sigma^2 (X^TX)^{-1}$.
\label{prop:Var(beta_S) Gaussian}
\end{theorem}
\begin{proof}
By the law of total variance, 
\begin{align*}
\text{Var}(\hat{\beta}_S) & = \text{Var}_Y(\E(\hat{\beta}_S|Y)) + \E_Y(\text{Var}(\hat{\beta}_S|Y)) \\
\end{align*}
Note that,
\[
\hat{\beta}_S = \left(X^TS^TSX\right)^{-1}X^TS^TS(X\hat{\beta}+e) = \hat{\beta}+ \left(X^TS^TSX\right)^{-1}X^TS^TSe,
\]
where $e = (I- X(X^TX)^{-1}X^T)Y$, which has mean $0$ and is independent of $S$ and $X\hat{\beta}$. Further, $X^Te=0$. We now derive the mean and variance of $\hat{\beta}_S$ conditioned on $Y$. 
Define $U := SX$ and $v := Se$. Since $X^Te =0$, and the entries of $S$ are Gaussian random variables, $U$ and $v$ are independent since correlation of each column of $U$ with $v$ is $0$ and $U$ and $v$ are jointly normal. Denote by $U^v$, the vectorization of $U$. We thus, get the distribution of $U$ and $v$ as:
\[
U^v \sim \N(0, I_m \otimes X^TX), \quad v \sim \N (0, \|e\|^2I_m)
\]
Therefore,
\[
E_S\left((U^TU)^{-1}U^Tv\right) = E_S((U^TU)^{-1}U^T)E_S(v) = 0
\]
We also have, letting $Y$ be fixed,
\begin{align*}
\Var_S(\hat{\beta}_S) & = \Var((U^TU)^{-1}U^Tv) \\
                      & = \Var_U(\E((U^TU)^{-1}U^Tv|U)) + \E_U(\Var((U^TU)^{-1}U^Tv|U))\\
                      & = \E_U(\|e\|^2 (U^TU)^{-1}),
\end{align*}
where the last equality follows from the fact that $v$ has variance $\|e\|^2I_m$. Now, $W := U^TU$ follows an $d\times d$ Wishart distribution with $m$ degrees of freedom and mean $X^TX$. It can be shown that, provided $m>d+1$,
\[
\E(W^{-1}) = \frac{1}{m-d-1} (X^TX)^{-1}
\]
Thus, we have the expression for the variance of $\hat{\beta}_S$,
\begin{align*}
\text{Var}(\hat{\beta}_S) & = \text{Var}_Y(\E(\hat{\beta}_S|Y)) + \E_Y(\text{Var}(\hat{\beta}_S|Y)) \\
                          & = \Var_Y(\hat{\beta}+0) + \E_Y(\|e\|^2\frac{1}{m-d-1}(X^TX)^{-1})\\
                          & = \sigma^2 (X^TX)^{-1} + \frac{n-d}{m-d-1}\sigma^2 (X^TX)^{-1}\\
                          & = \frac{n+m-2d-1}{m-d-1} \sigma^2 (X^TX)^{-1}
\end{align*}
\end{proof}
In this case, we see that the variance of the estimator is inflated by a factor of the order of about $\frac{n}{m}$, which is substantial. Since both $\hat{\beta}$ and $\hat{\beta}_S$ are unbiased their variance is also their Mean squared error. From the above results, we have that the relative efficiency of $\hat{\beta}$ with respect to $\hat{\beta}_S$ is about $\frac{n}{m}$, i.e, for $\hat{\beta}_S$ to be as efficient at $\hat{\beta}$ at the current sample size it has to have a sample size about $\frac{n}{m}$ times larger than what is available.

\subsection{Asymptotic Distributions for Hadamard and Clarkson-Woodruff Sketches}
The JL Transforms we consider in the paper to get an improvement in the computational complexity of solving the $L_2$ regression problem are the Fast JL or Hadamard Transform (\cite{ailon2009fast}) and sparse embeddings such as the Clarkson-Woodruff transform (\cite{clarkson2017lowrank}). For these sketches, it is often mathematically intractable to obtain a finite sample distribution of the sketched dataset and hence of the estimate $\hat{\beta}_S$, one major reason being dependence between the entries of the matrix. Thus asymptotic results are useful here. \cite{Ahfock2021} states a sketching CLT under the an assumption which holds for both the Hadamard and Clarkson-Woodruff transforms.
We define a sequence of data matrices and responses $X_{(n)} \in \mathbb{R}^{n \times d},y_{(n)}\in \mathbb{R}^n$. Define $A_{(n)}:=[y_{(n)} \quad X_{(n)}]$ as the source dataset. Let $\tilde{A} = [\tilde{y} \quad \tilde{X}]$ denote the sketched dataset, $SA_{(n)}$ for the sketching matrix $S$.
\begin{assumption}
Let the singular value decomposition of the dataset be given by $A_{(n)} = U_{(n)}D_{(n)}V_{(n)}^T$. Let $u_{n(i)}^T$ denote the $i$th row of $U_{(n)}$. Assume that 
\begin{align*}
    \lim_{n \rightarrow\infty} \max_{i \in [n]}\|u_{n(i)}\|^2 = 0.
\end{align*}
This is also known as the maximum leverage score.
\label{assumption:A1 for Asymptotic Sketched Distribution}
\end{assumption}
This gives us Theorem 3 of \cite{Ahfock2021}.

\begin{proposition}
\label{thm: Sketching CLT}
Consider a fixed sequence of arbitrary $n \times (d+1)$ data matrices $A_{(n)}$, where $d$ is fixed. Let $A_{(n)} = U_{(n)}D_{(n)}V_{(n)}^T$ represent the singular value decomposition of $A_{(n)}$. Let $S$ be a $m \times n$ Hadamard or Clarkson-Woodruff sketching matrix where $m$ is also fixed. Suppose that Assumption \ref{assumption:A1 for Asymptotic Sketched Distribution} on the maximum leverage score is satisfied. Then as $n$ tends to infinity with $m$ and $d$ fixed, we have the following convergence in distribution:
\[
[\tilde{A}V_{(n)}D_{(n)}^{-1}| A_{(n)}] \to \mathcal{MN}(0, I_m, I_{d+1}/m)
\]
\end{proposition}
One further assumption is required to obtain asymptotic properties, which is standard  and ensures stability of the $\hat{\beta}, \hat{\sigma}^2_n$ and hence $\hat{\beta}_S$ estimates.
\begin{assumption}
\[
\lim_{n \to \infty} n^{-1} 
\begin{bmatrix} 
y_{(n)}^T y_{(n)} & y_{(n)}^T X_{(n)} \\ 
X_{(n)}^T y_{(n)} & X_{(n)}^T X_{(n)} 
\end{bmatrix} 
= Q
\]
for some positive-definite matrix $Q$.
\label{assumption: A2 for Asymptotic Sketched Distribution}
\end{assumption} 
Under both these assumptions, it is possible to prove asymptotic normality for $\hat{\beta}_S$.
\begin{proposition}
\label{thm: Sketched Estimator Asymptotic Distribution}
Suppose that Assumptions 1 and 2 hold, $m \ge d$, and $\beta_S$ is computed using a Hadamard or Clarkson-Woodruff sketch. Let $(\tilde{X}^T \tilde{X})^+$ denote the Moore-Penrose pseudo-inverse of $(\tilde{X}^T \tilde{X})$. Let
\[
\tilde{V}_{(n)} = \frac{RSS_F^{(n)}}{m}(\tilde{X}^T \tilde{X})^+ \quad \text{and} \quad V_{(n)} = \frac{RSS_F^{(n)}}{m-d+1}(X_{(n)}^T X_{(n)})^{-1}
\]
Then as $n \to \infty$, convergence in distribution holds for
\begin{itemize}
    \item[(i)] $[V_{(n)}^{-1/2}(\beta_S - \beta_F^{(n)}) | A_{(n)}] \to \text{Student}(0, I_d, m-d+1)$
    \item[(ii)] $[\tilde{V}_{(n)}^{-1/2}(\beta_S - \beta_F^{(n)}) | A_{(n)}] \to \mathcal{N}(0, I_d)$
\end{itemize}
\end{proposition}
Proposition \ref{thm: Sketching CLT} shows that the sketched dataset behaves approximately like the the Gaussian sketched dataset as $n \rightarrow \infty$. Since $\hat{\beta}_S$ is a continuous function of the sketched data matrix, the continuous mapping theorem gives us that conditioned on $A_{(n)}$, the sketched estimator is also asymptotically normal. Under the Assumption 2, Proposition \ref{thm: Sketched Estimator Asymptotic Distribution} establishes this asymptotic distribution. Of $V_{(n)}$ and $\tilde{V}_{(n)}$, $\tilde{V}_{(n)}$ is easier to compute as we have seen. 

While the classical OLS estimator achieves an asymptotic variance of $\sigma^2(X^T X)^{-1}$, dimensionality reduction injects substantial structural noise. For a Gaussian sketch of dimension $m \ll n$, the finite-sample variance strictly inflates to $\frac{n+m-2d-1}{m-d-1}\sigma^2(X^T X)^{-1}$. For Hadamard and Clarkson-Woodruff sketches, exact finite-sample moments may diverge absent strict uniform integrability. However, their asymptotic equivalence in distribution to the Gaussian sketch (Proposition \ref{thm: Sketched Estimator Asymptotic Distribution}) yields an asymptotic variance of approximately $\frac{n-d}{m-d+1}\sigma^2(X^T X)^{-1}$, scaling the baseline standard error by a factor of roughly $\sqrt{n/m}$.

Furthermore, sketching inflicts a secondary distributional penalty on inference. The scaled sketched estimator converges to a heavier-tailed Student-t distribution, $[V_{(n)}^{-1/2}(\beta_S - \beta_F^{(n)})\vert A_{(n)}] \xrightarrow{d} \text{Student}(0, I_d, m-d+1)$. Consequently, when constructing confidence sets: intervals are widened by both the inflated standard error and the larger critical values demanded by the restricted $m-d+1$ degrees of freedom. 

\subsection{Efficiency of Sketched Estimators}

We have noted that standard inferential procedures produce much worse results for the sketched estimator. Another perspective on the statistical properties of the sketched estimator is on it's efficiency. In this regard, \cite{DobribanLiu2019} and \cite{RaskuttiMahoney2016} compare the sketched estimator against the OLS estimator broadly under the following efficiency metrics:

\begin{itemize}
\item \textbf{Variance Efficiency (VE):} $VE(\hat{\beta}_S, \hat{\beta}) = \frac{\mathbb{E}\Vert{}\hat{\beta}_S - \beta\Vert{}^2}{\mathbb{E}\Vert{}\hat{\beta} - \beta\Vert{}^2}$, which measures the inflation in parameter estimation error.
\item \textbf{Prediction Efficiency (PE):} $PE(\hat{\beta}_S, \hat{\beta}) = \frac{\mathbb{E}\Vert{}X\hat{\beta}_S - X\beta\Vert{}^2}{\mathbb{E}\Vert{}X\hat{\beta} - X\beta\Vert{}^2}$, which captures the loss of accuracy in predicting the regression function.
\item \textbf{Out-of-sample Efficiency (OE):} $OE(\hat{\beta}_S, \hat{\beta}) = \frac{\mathbb{E}(x_t^\top \hat{\beta}_S - y_t)^2}{\mathbb{E}(x_t^\top \hat{\beta} - y_t)^2}$, evaluated on an independent test point $(x_t, y_t)$.
\item \textbf{Residual Efficiency (RE):} $C_{RE}(S) = \frac{\mathbb{E} \left[ \Vert{} Y - X \beta_S \Vert{}_2^2 \right]}{\mathbb{E} \left[ \Vert{} Y - X \beta_{OLS} \Vert{}_2^2 \right]}$, evaluated for a fixed sketching matrix $S$.
\end{itemize}

\cite{RaskuttiMahoney2016} provides upper and lower bounds for the prediction and residual efficiency, while \cite{DobribanLiu2019} evaluates asymptotic values of the various relative efficiencies under different growth rates of the original and reduced number of datapoints $m,n$ and the dimension of the data $d$. We state these results for the Hadamard and IID sketches for completeness.

\begin{proposition}[Universality for i.i.d. Projection, Theorem 2.2 \cite{DobribanLiu2019}]
Suppose $S$ has i.i.d. entries of zero mean and finite fourth moment, and $X$ is a deterministic matrix with singular values uniformly bounded away from zero and infinity. As $n \to \infty$ with $d/n \to \gamma$ and $m/n \to \xi$, the efficiencies converge to:
\begin{equation*}
\lim_{n \to \infty} VE(\hat{\beta}_S, \hat{\beta}) = \lim_{n \to \infty} PE(\hat{\beta}_S, \hat{\beta}) = 1 + \frac{1 - \gamma}{\xi - \gamma}
\end{equation*}
\label{prop:Liu thm 2.2}
\end{proposition}

\begin{proposition}[Randomized Hadamard Transform, Theorem 2.4 \cite{DobribanLiu2019}]
Let $S$ be an $m \times n$ randomized Hadamard matrix (SRHT). Under the same assumptions on $X$ as \ref{prop:Liu thm 2.2}, the limits of the variance and prediction efficiencies are equivalent to those of the Haar projection:
\begin{equation}
\lim_{n \to \infty} VE(\hat{\beta}_S, \hat{\beta}) = \lim_{n \to \infty} PE(\hat{\beta}_S, \hat{\beta}) = \frac{1 - \gamma}{\xi - \gamma}
\end{equation}
\label{prop:Liu thm 2.4}
\end{proposition}

These indicate that asymptotically, the sketched estimator has lower predictive efficiency than that of the $\hat{\beta}$ estimator.
The following result --- Theorem 1 of \cite{PilanciWainwright2016}, proves a non-asymptotic probabilistic lower bound on the prediction efficiency of the  sketched estimator which proving the suboptimality of the sketched estimator when the errors in the model $\epsilon$ follow iid normal distributions.

\begin{proposition}[Lower Bound on Prediction Efficiency, Theorem 1 \cite{PilanciWainwright2016}]
For any sketching matrix satisfying $\Vert{}\mathbb{E}[S^\top(SS^\top)^{-1}S]\Vert{}_{op} \le \eta\frac{m}{n}$, any estimator based on $(SX, SY)$ satisfies the lower bound with probability greater than $1/2$:
\begin{equation*}
PE(\hat{\beta}_S,\hat{\beta}) \ge \frac{n}{128\eta m}
\end{equation*}
\label{prop:PE Lower Bound}
\end{proposition}
\section{Conclusion}
 Through our analysis, we have found that for subspace embedding based on JL lemma and its modified versions to hold for linear regression, we require strict conditions on the data, which is not always practically feasible. We have also converted some well known theorems in this field into concrete forms which give precise conditions for subspace embedding methods to hold.
 Theorem \ref{JL subspace embedding} gives a sufficient condition for Johnson-Lindenstrauss based subspace embedding to hold. A similar sufficient condition is obtained for Fast JL based subspace embedding in Theorem \ref{FastJL thm}. 
 
   We have also compared the time complexity for Fast JL and Sparse JL embeddings to regular linear regression. For sparse embeddings like OSNAP we see that the lower bound on the reduced dimension is very stringent and that leads to no computational benefit in practice for small values of $\epsilon,\delta$. We can also infer by Theorem \ref{sparsity 1 lower bound} that irrespective of the embedding the stringent inequality on the reduced dimension cannot be made much better for small values of $\epsilon,\delta$. In the later section, we also compare these methods to solutions using iterative methods like LSQR and LSMR where we find that for regression based on sketching to improve time complexity we need a large number of data points . While classical LSMR is a robust iterative solver for ill-conditioned matrices, applying sketching inflates the condiiton number of the data matrix by a factor of $\frac{1+\epsilon}{1-\epsilon}$. We see that, for the sketched LSMR approach to provide computational advantage over the classical method, the original number of data points should be large compared to the reduced dimension. Our analysis focuses on the Sketch-and-Solve framework; a similar analysis of the Precondition-and-Solve method can be a direction for future work.
   
   In addition to sufficient conditions, we establish necessary bounds on the reduced dimension in Theorem \ref{thm:Concrete reduced dimension bounds} and derive a lower bound for embeddings with sparsity one in Theorem \ref{sparsity 1 lower bound}. Together, these results provide a more concrete picture of the dimensional requirements and limitations of JL-based methods in linear regression.
   
   Although sharper proof techniques may improve the constants in our bounds, the dependence of the reduced dimension on the target accuracy remains an important consideration. In particular, the inverse-quadratic dependence on the accuracy parameter ($\epsilon$) can make the theoretical guarantees restrictive when high accuracy is required or when the available data dimension is not sufficiently large. This highlights the importance of examining not only the asymptotic guarantees of subspace embedding methods but also the parameter regimes in which they offer practical computational advantages.

   Furthermore, dimensionality reduction imposes severe statistical penalties on the resulting estimators. Although sketched regression estimators remain structurally unbiased, their variance is substantially inflated; a Gaussian sketch, for example, inflates the variance by a factor of $\frac{n+m-2d-1}{m-d-1}$. The asymptotic distributions of the Clarkson-Woodruff and Hadamard estimators are seen to also have an inflated asymptotic variance. Additionally, sketched estimators exhibit lower prediction efficiency than standard Ordinary Least Squares (OLS) estimators and converge asymptotically to heavier-tailed Student-t distributions, which ultimately widens confidence intervals by large factors, compromising inference. 

An interesting direction for future work is to investigate whether analogous dimensional limitations arise in other applications of JL-based methods, such as matrix multiplication (\cite{sarlos}) and $k$-nearest-neighbour algorithms (\cite{ailon2009fast}). Such an investigation would require application-specific analyses of the corresponding error guarantees and computational trade-offs.

\section*{Acknowledgment}
 We are extremely grateful to Palash Sarkar, Indian Statistical Institute, Kolkata, for his constant guidance and support throughout the course of the work.
\bibliographystyle{plainnat}
\bibliography{references}
\end{document}